\documentclass[epsfig,graphics,usenatbib]{mn2e}
\usepackage{times}
\usepackage{graphicx}
\usepackage{psfrag}
\usepackage{amsmath,amssymb}
\usepackage{threeparttable}
\begin{document}
\newcommand{\Zsolar}{\mbox{$\,\rm Z_{\odot}$}}
\newcommand{\Msolar}{\mbox{$\,\rm M_{\odot}$}}
\newcommand{\Lsolar}{\mbox{$\,\rm L_{\odot}$}}
\newcommand{\xs}{$\chi^{2}$}
\newcommand{\ls}{{\tiny \( \stackrel{<}{\sim}\)}}
\newcommand{\gs}{{\tiny \( \stackrel{>}{\sim}\)}}
\newcommand{\asec}{$^{\prime\prime}$}
\newcommand{\amin}{$^{\prime}$}
\newcommand{\bfs}[1]{\mbox{\boldmath $#1$}}
\newcommand{\argmax}[1]
           { {\renewcommand{\arraystretch}{0.5}
              \begin{array}{ccc} ~ \\ \mbox{\rm argmax} \\ {\scriptstyle #1}
              \end{array} } }
\newcommand{\bd}[1]{\boldsymbol{#1}}
\newcommand{\ny}{\left\{}
\newcommand{\lln}{\langle}
\newcommand{\rrn}{\rangle}
\newcommand{\zr}{\right\}}
\newcommand{\cc}{\mathcal N} 
\newcommand{\etal}{{\it et al.}} 
\newcommand{\eg}{{\it e.g.}}
\newcommand{\ie}{{\it i.e.}}
\newcommand{\ee}{\operatorname{erfc}}
\newcommand{\pa}{\operatorname{pa}}
\newcommand{\Var}[1]{\operatorname{Var}\{#1\}}
\newcommand{\RR}{\mathcal R}
\title[Young stellar populations in early-type galaxies]{A data-driven Bayesian approach for finding young stellar populations in early-type galaxies from their UV-optical spectra}
\author[L.A. Nolan \etal]
{Louisa A. Nolan$^{1}$, Markus O. Harva$^{2}$, Ata Kab\'{a}n$^{3}$ $\&$ Somak Raychaudhury$^{1}$\\
$^{1}$School of Physics and Astronomy, University of Birmingham, Birmingham, B15 2TT, UK\\
$^{2}$Laboratory of Computer and Information Science, Helsinki University of Technology, Helsinki, P.O. Box 5400, FI-02015 HUT, Finland\\
$^{3}$School of Computer Science, University of Birmingham, Birmingham, B15 2TT, UK}

\date{Submitted for publication in MNRAS}

\maketitle
  
\begin{abstract}
Efficient predictive models and data analysis techniques for the
analysis of photometric and spectroscopic observations of galaxies are
not only desirable, but required, in view of the overwhelming
quantities of data becoming available. We present the results of a
novel application of Bayesian latent variable modelling techniques,
where we have formulated a data-driven algorithm that allows one to
explore the stellar populations of a large sample of galaxies from
their spectra, without the application of detailed physical models.
Our only assumption is that the galaxy spectrum can be expressed as a
linear superposition of a small number of independent factors, each a
spectrum of a stellar sub-population that cannot be individually
observed.  A probabilistic latent variable architecture that
explicitly encodes this assumption is then formulated, and a rigorous
Bayesian methodology is employed for solving the inverse modelling
problem from the available data.  A powerful aspect of this method is
that it formulates a density model of the spectra, based on which we
can handle observational errors.  Further, we can recover missing data
both from the original set of spectra which might have incomplete
spectral coverage of each galaxy, or from previously unseen spectra of
the same kind.

We apply this method to a sample of 21 ultraviolet-to-optical spectra
of well-studied early-type galaxies, for which we also derive detailed
physical models of star formation history (i.e. age, metallicity and
relative mass fraction of the component stellar populations). We also
apply it to synthetic spectra made up of two stellar
populations, spanning a large range of parameters. We apply four
different data models, starting from a formulation of principal
components analysis (PCA), which has been widely used. We explore
alternative factor models, relaxing the physically unrealistic
assumption of Gaussian factors, as well as constraining the possibility of
negative flux values that are allowed in PCA, and show that other
models perform equally well or better, while yielding more physically
acceptable results. In particular, the more physically motivated assumptions of 
our rectified factor analysis enable it to perform better than PCA, and to recover physically meaningful results.

We find that our data-driven Bayesian modelling allows us to identify
those early-type galaxies that contain a significant stellar
population that is \ls\ 1 Gyr old. This experiment also concludes that
our sample of early-type spectra showed no evidence of more than two
major stellar populations differing significantly in age and
metallicity.  This method will help us to search for such young
populations in a large ensemble of spectra of early-type galaxies,
without fitting detailed models, and thereby to study the underlying
physical processes governing the formation and evolution of early-type
galaxies, particularly those leading to the suppression of star
formation in dense environments. In particular, this method would be a
very useful tool for automatically discovering various interesting
sub-classes of galaxies, for example post-starburst, or E+A galaxies.

\end{abstract}

\begin{keywords}
galaxies: elliptical and lenticular, cD -- galaxies: stellar content
-- methods: data analysis
\end{keywords}

\section{Introduction}

The determination of the star formation history of early-type galaxies
has important implications for the still-controversial issue of their
formation and evolution. Are the bulk of their stars assembled at high
redshift, followed by predominantly passive evolution, or are they
formed from low-redshift disk-disk merging? Or, as it seems likely
from the variety of detailed elliptical morphology which is being
revealed via projects such as those involving
integral field spectrographs such as SAURON, does the
class of ellipticals contain sub-classes, and are these dependent on
the method of their formation, or on their local environment
\citep{sauron01,ems04}? We would
like to know how these sub-classes depend on local environment. In
particular, we wish to know what mechanisms are responsible for the
now well-known morphology-density relationship
\cite[\eg][]{dressler80}, which shows that the proportion of early-type
galaxies increases as the proportion of late-type galaxies decreases
in regions with higher galaxy density. Furthermore, the recent interest
in E+A galaxies \cite[\eg][]{tran03}, in which the signature of a recent
(\ls\ 1 Gyr) starburst is seen, but no significant
ongoing star formation, suggests that the relation between the evolution
of these galaxies and their
constituent stellar populations is not well-understood.

However, until recently, only small ensembles of galaxy spectra,
spanning a large useful range, were available and the analysis of a
large statistical sample of stellar populations of galaxies was not
possible. With the recently-completed 2dF Galaxy Redshift Survey
(2dFGRS) and the ongoing Sloan Digital Sky Survey (SDSS)
or 6dF Galaxy Survey (6dFGS), a few
million galaxy spectra will soon be available. The detailed physical
modelling of stellar populations is numerically expensive, so the
timely extraction of useful knowledge, such as the characteristics of
the star formation history of galaxies, from these data will largely
depend on developing appropriate data analysis tools that are able to
complement more specialised astrophysical studies, in particular by
automating parts of the analyses. Some work has already been carried out using 
an approach which incorporates assumptions from physical models to a greater or 
lesser extent \citep[\eg][]{hjl00,k03,f05,oc05a,oc05b,sol05}.

With the increase of data availability, significant research has
recently been focusing on the use of data-driven methods of
astrophysical data analysis. Some of these attempt to analyse galaxy
properties from spectra, independently from any physical model. The
usefulness of the application of Principal Component Analysis (PCA), a
classical multivariate statistical analysis technique, has been
exploited and demonstrated in various scenarios
\citep[see, \eg][]{connolly95,folkes96,ronen99,madg03a,madg03b}.
Indeed, PCA is an
efficient tool for data compression, and as such, suitable for
providing a compressed representation for subsequent physical
analysis. In a different line of research, the automated partitioning
of spectra based on their morphological shapes \citep{slonim01}
has also been explored with reasonable success.

Here, we explore the application of a similar technique to a different
problem, namely that of the data-driven identification of the spectra
of distinct stellar sub-populations in the spectrum of a galaxy. Our
only assumption is that the galaxy spectrum can be expressed as a
linear superposition of a small number of independent factors, each a
spectrum of a stellar sub-population that cannot be individually
observed. We would like to implement an optimal feature transformation
that would reveal interpretable and meaningful structural properties
(factors) from a large sample of galaxy spectra.  This would allow
fast, automated extraction of some key physical properties from
potentially very large data sets of the kind the Virtual Observatory
would offer.

Preliminary results based on a simple non-orthogonal projection method
have indicated that an identification of interpretable component
spectra may be achievable in a purely data-driven manner
\citep{knr05}.  In this work, we follow a more rigorous methodology,
explicitly encoding the above hypothesis into appropriate
probabilistic data models. We then employ a Bayesian
formalism to solve the inverse modelling problem, as well as to
objectively assess to what extent such a hypothesis is supported by
the data, in terms of both physical interpretability and 
the ability for predictive
generalisation to previously unseen data of the same type.

The Bayesian framework allows us to automate several functional
requirements within a single statistically sound formal framework,
where all modelling assumptions may be explicitly encoded. Tasks
ranging from inference and prediction to inverse modelling and data
explanation, data compression, as well as model comparison are all
accomplished in a flexible manner, even in conditions of missing
values and measurement uncertainties.

In this study we start from a set of synthetic spectra of stellar
populations of known age and metallicity from the single stellar
population models of \citet{jimenez04}. We also have assembled a set
of observed ultraviolet (UV)-optical spectra of early-type galaxies, to
which we have fit two component stellar population models,
and thus we have some knowledge of their star formation history
(i.e. ages, metallicities and mass fractions of the component stellar
populations).  We then perform our data-driven Bayesian analysis on
both synthetic and observed spectra, and compare the derived
parameters with those obtained from detailed astrophysical modelling.

In \S2, we describe the set of 21 observed early-type galaxy
spectra that we analyse here, and  describe the synthetic
stellar population spectra. \S3 details the Bayesian modelling of the
data. Our results are presented in and discussed in \S4. In
\S5  we discuss the significance of this work on characterising
E+A galaxies.  Our conclusions are outlined in \S6. Appendix A
gives the detailed mathematical framework required to implement our
methodology, specifically for the data model estimation of the variational
Bayesian rectified factor analysis --- a flexible data model for the 
factorisation of positive data that we developed.

\section{The Data}\label{obs}

\subsection{Synthetic stellar population models}\label{synthetic}

We perform the linear independent transformation analyses on a
set of two-component model stellar population spectra. We use the
single stellar population models of \citet{jimenez04}, which have a Salpeter initial mass function, across the
same wavelength range (2500$-$8000 \AA) as the observed spectra, and
create a two-component model flux, $ F_{2pop,\lambda}$. This is
defined as
\begin{equation}
 F_{2pop,\lambda}  \propto 
 \left( m_{i} f_{Z_{i},\lambda,t_{i}} + m_{j} f_{Z_{j},\lambda,t_{j}} 
\right) 
\label{modflux} 
\end{equation}
where $F_{2pop,\lambda,t}$ is the model flux per unit wavelength in
the bin centred on wavelength $\lambda$, which is the sum of the two
single-metallicity, single-age model fluxes 
$f_{Z_{i},\lambda,t_{i}}$
and $f_{Z_{j},\lambda,t_{j}}$, which have metallicities and ages,
$Z_{i}, t_{i}$ and $Z_{j}, t_{j}$ respectively.  Here $m_{i}, m_{j}$ are
the fractional contributions (by stellar mass) of each component to
the total model population. We set
$m_{i} + m_{j} = 1$, where $m_{i}$ can take
values 0.25 or 0.5. Ages assigned for $t_{i}, t_{j}$ are 0.03, 1, 3, 7
and 12 Gyr, and the metallicities are 0.004, 0.01, 0.02, 0.03 and
0.05, making a total of 50 two-component model spectra. 

\subsection{Observed early-type galaxy spectra}

\begin{figure}
\begin{center}
\includegraphics[width=21.0cm,angle=-90]{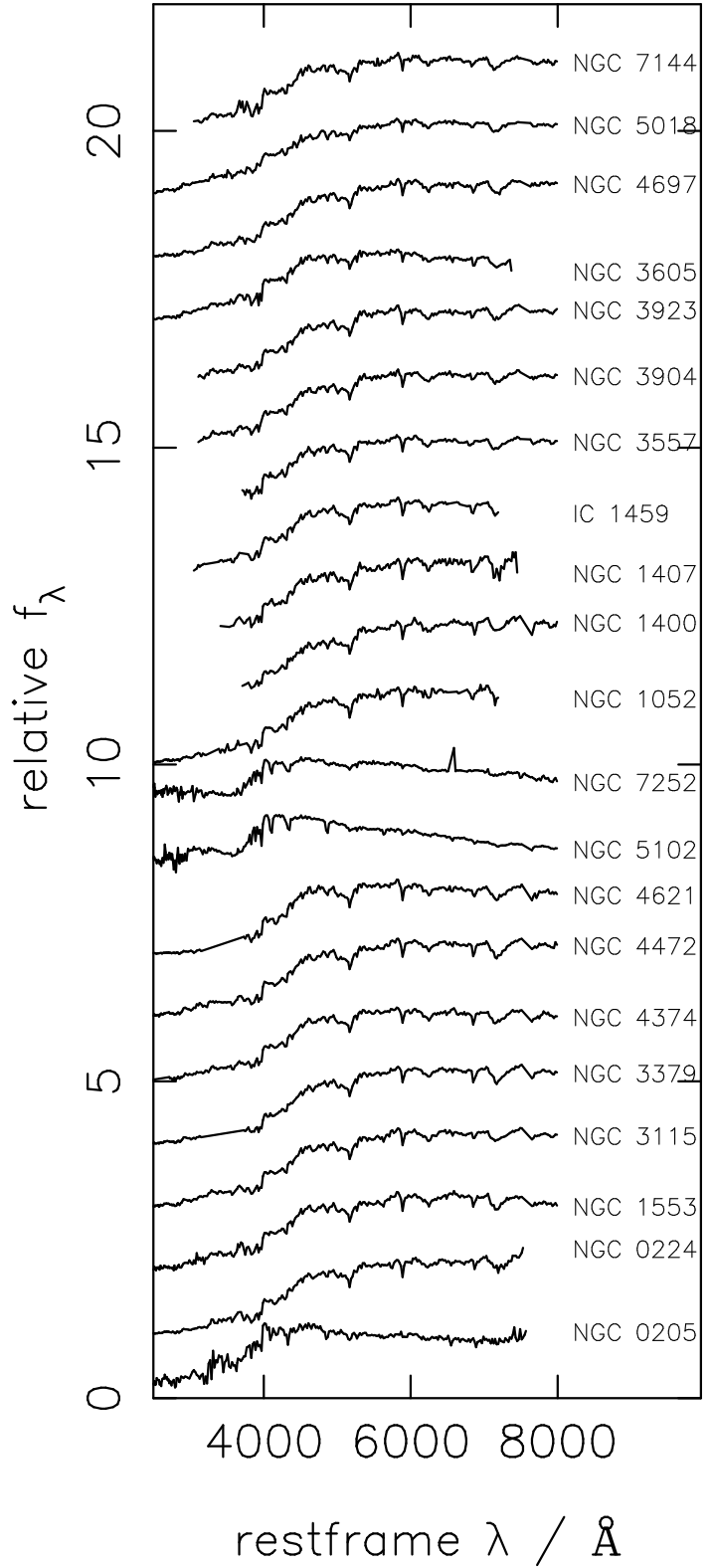}
\caption{Rest-frame spectra of the 21 galaxies in our sample. 
See Table \ref{data} and \S\ref{obs} for details. The dotted lines
mark some of the absorption features in the spectra which are
typically strong in young stellar populations, and the dashed lines
mark some of the absorption features which are typically strong in
old, metal-rich stellar populations. From left to right, the
absorption line species are: MgII (2799 \AA), H$\varepsilon$ (3970
\AA), H$\delta$ (4102 \AA), H$\gamma$ (4340 \AA), H$\beta$ (4861
\AA), Mgb (5175 \AA), NaD (5893 \AA), H$\alpha$ (6563 \AA), TiO
(7126 \AA). The spectra are arbitrarily shifted along the flux axis
for the sake of clarity.}
\label{spec}
\end{center}
\end{figure}

We have also compiled a set of twenty-one early-type galaxy 
spectra from the archives for this study, 
across the wavelength range 2500$-$8000
\AA, although not all objects have data covering the complete
range. The spectra of the galaxies in the sample are shown in 
Fig.~\ref{spec}.  
As the galaxy sample was assembled from those nearby
early-type galaxies with both UV and optical spectra, from various
archives, it does not represent a statistical sample. The sources for
the data are listed in Table \ref{data}, together with the telescopes
with which the optical spectra were observed. The UV spectra were
observed using the {\it International Ultraviolet Explorer} (IUE),
except for NGC 3605 and NGC 5018, which were observed with the {\it
Faint Object Spectrograph} (FOS) on the {\it Hubble Space Telescope}
(HST).

Since these spectra are compiled from sources with varying spectral
coverage, the resulting data matrix has some missing values.  The
regions of incomplete wavelength coverage are obvious in Fig.~\ref{spec}.  It is one of the strengths of our analysis that these
missing data regions can be recovered; this is described in \S\S
\ref{missdat}, \ref{missdat2}, \ref{missdat3}. Another of the advantages of the
probabilistic framework adopted here is that, in contrast with classic
non-probabilistic methods, the observational errors on each data point
can be taken into account (\S\ref{errors}).

As the data matrix on which the analysis is performed must have the
same wavelength binning for all the galaxies, the spectra are 
de-redshifted and binned
to the same wavelength resolution as the synthetic stellar population
spectra ($\sim$20 \AA; the observed spectra have wavelength
resolution which varies from $\sim$5-18 \AA). For each galaxy, the
various spectral sections, taken in turn, are normalised to unity in
the overlap regions, and spliced, to create a single continuous
spectrum for each object. Although the different spectral pieces for any one object were observed separately, and with different telescopes, it should be noted that the apertures used were similar, and therefore each section probes a similar part of each galaxy, and the shape of the continua in the overlap regions agree well.

\begin{table}
\begin{threeparttable}
\begin{raggedright}
\caption{Table of references and optical telescopes for the observed early-type galaxy spectra. }
\label{data}
\begin{tabular}{lll}
\hline      
{Object} & {Reference} & {Optical telescope} \\
\hline
NGC 0205 &  S-BCK  & 0.9m, Kitt Peak National Observatory  \\
NGC 0224 &  S-BCK  & 0.9m, Kitt Peak National Observatory  \\
NGC 1052 &  S-BCK  & 0.9m, Kitt Peak National Observatory  \\
NGC 1400 &  Q      & ESO 1.52m, La Silla  \\
NGC 1407 &  S-BCK  & 0.9m, Kitt Peak National Observatory  \\
IC 1459	 &  R,Q    & ESO 2.2m, 1.52m, La Silla  \\
NGC 1553 &  R,Q    & ESO 2.2m, 1.52m, La Silla  \\
NGC 3115 &  R,Q,B  & ESO 2.2m, 1.52m, La Silla  \\
NGC 3379 &  Q      & ESO 1.52m, La Silla  \\
NGC 3557 &  Q,B    & ESO 2.2m, 1.52m, La Silla  \\
NGC 3605 &  P,J    & F.L. Whipple Observatory, 1.5m  \\
NGC 3904 &  R,Q,B  & ESO 2.2m, 1.52m, La Silla  \\
NGC 3923 &  R,Q,B  & ESO 2.2m, 1.52m, La Silla  \\
NGC 4374 &  R,Q    & ESO 2.2m, 1.52m, La Silla  \\
NGC 4472 &  R,Q    & ESO 2.2m, 1.52m, La Silla  \\
NGC 4621 &  Q      & ESO 1.52m, La Silla  \\
NGC 4697 &  R,Q,B  & ESO 2.2m, 1.52m, La Silla  \\
NGC 5018 &  P,R,Q,B& ESO 2.2m, 1.52m, La Silla  \\
NGC 5102 &  R,Q,B  & ESO 2.2m, 1.52m, La Silla  \\
NGC 7144 &  R Q B  & ESO 2.2m, 1.52m, La Silla  \\
NGC 7252 &  V      & 1.93m, Haute Provence Observatory  \\
\hline
\end{tabular}
\begin{tablenotes}
\item[ ] The UV spectra were observed with the IUE, except for NGC 3605 and NGC 5018, which were observed using FOS on the HST.
\item[ ]{\bf Sources:}
    J: \citet{jansen00};
    P: \citet{ponder98};
    S-BCK: \citet{sbck98};
    R: Bica \& Alloin, unpublished;
    Q: \citet{bc87a};
    B: \citet{bc87b};
    V: Bica \& Alloin, unpublished, from 
   {\sl ftp://cdsarc.u-strasbg.fr/cats/III/219/}
\end{tablenotes}
\end{raggedright}
\end{threeparttable}
\end{table}

\section{Data modelling}

Models are simplifications of reality representing relevant aspects
while glossing over irrelevant details.  Data models are constructed
from experimental evidence (data) and available prior knowledge.  Once
a data model is estimated (which can be done off-line), it offers a
tool for making inferences and predictions in a consistent framework.

In Bayesian data modelling, all information is encoded in probability
distributions \citep{bs94}. The joint probability distribution of the
observed data and other unobservable variables of interest (such as
factors or component stellar populations of the representation as well
as the parameters of the mixing process, etc.) need to be designed.
The unobservable variables in the model are often referred to as
hidden or latent variables.

\subsection{Designing and building an appropriate latent variable architecture}

\subsubsection{The independent factor representation model of the spectrum}
The spectrum of a galaxy or of a stellar sub-population is
essentially a vector, over a binned wavelength range, representing the
flux per unit wavelength bin. Consider a set of $N$ spectra of
early-type galaxies, each measured at $T$ different wavelength
bins. We denote by $\bd{x}_t
\in \RR^N$ the vector formed by the flux values at the wavelength bin
indexed by $t$, as measured across the $N$ spectra. The $N \times T$
matrix formed by these vectors may be referred to as $\bd{X}$ and
single elements of this matrix will be denoted as $x_{nt}$. Similar
notational convention will also apply to other variables used.

As stated above, the hypothesis of our model is that each of the $N$
observations can be represented as a superposition of $K<N$ latent
underlying component spectra or factors $f(\bd{s}_t) \in \RR^{K}$,
that are not observable by direct measurements but only through an
unknown linear mapping $\bd{A} \in \RR^{N \times K}$.  Formally, this
can be summarised as follows:
\begin{equation}
\bd{x}_t = \bd{A}f(\bd{s}_t)+\bd{\epsilon}.
\label{model}
\end{equation}
In (\ref{model}), the first term of the right is the so-called
systematic component and $\bd{\epsilon}$ is the noise term or the
stochastic component of this model, both of which will be further
detailed below. The noise term $\bd{\epsilon}$ is assumed to be a
zero-mean independent and identically distributed Gaussian, with
variance $e^{-v_{x_n}}$.  The diagonal structure of the covariance
accounts for the notion that all dependencies that exist in $\bd{x}_t$
should be explained by the underlying hidden factors.

The $K$ components are assumed to be statistically independent, which
is a standard assumption of independent factor models.
The role of the function $f$ here is a technically convenient way of
imposing certain constraints on the components $f(\bd{s})$ in order to
facilitate the estimation of \emph{interpretable} factors.  This will
be detailed in subsection 3.1.4.

As in nearly all data modelling applications of factor models, we hope
that the feature transformation produced by the mapping $\bd{A}$ will
reveal important structural characteristics of the collection of
spectra. It is part of our data modelling design to investigate
various ways of facilitating this, as will be detailed in \S3.1.4.

\subsubsection{Dealing with measurement errors 
in the data model}\label{errors}

Classical non-probabilistic data analysis tools do not offer the
flexibility required for taking into consideration the uncertainty
that is associated with all real measurements.  It is thus an
important practical advantage of the probabilistic framework we adopt
here that it allows us to take such measurement errors into account as
an integral part of the data model.  This is achieved simply by making
the `clean' vectors $\bd{x}_t$ in (\ref{model})
 become hidden variables of the
additional error model
\begin{equation}
y_{nt} = x_{nt} + e_{nt},
\end{equation}
where $e_{nt}$ is a zero-mean Gaussian noise term with known standard
deviations $\sigma_{nt}$, for each individual measurement
$n=1:N,t=1:T$.

\subsubsection{Dealing with missing values in the data model}\label{missdat}

Our probabilistic framework also allows us to treat missing values in
a principled and straightforward manner under the assumption
that they are missing at random
\citep{gj94,cls03}.
Splitting each datum vector into missing and observed parts, i.e.
$\bd{y}_t^T=(\bd{y}_t^{oT}, \bd{y}_t^{mT})$, the missing entries
simply marginalise out:
\begin{equation}
p(\bd{y}_t^o) = \int p(\bd{y}_t)d\bd{y}_t^m.
\end{equation}
The superscript `$o$' will be omitted hereafter
for simplicity, noting that unless otherwise stated, $\bd{y}_t$ will
refer to $\bd{y}_t^o$ and missing entries are discounted in all the
update equations that follow.

Having completed the general structural specification of the proposed
probabilistic model, in the following subsection, we will make more
specific assumptions about distributional form of the latent spectral
representation that emerges from such probabilistic models.

\begin{figure}
\centering
\psfrag{s_1}[cc][cc]{$s_{1}$}
\psfrag{s_K}[cc][cc]{$s_{K}$}
\psfrag{x_1}[cc][cc]{$x_{1}$}
\psfrag{x_N}[cc][cc]{$x_{N}$}
\psfrag{vx_1}[cc][cc]{$v_{x1}$}
\psfrag{vx_N}[cc][cc]{$v_{xN}$}
\psfrag{a_11}[cc][cc]{$a_{11}$}
\psfrag{a_1K}[cc][cc]{$a_{1K}$}
\psfrag{a_N1}[cc][cc]{$a_{N1}$}
\psfrag{a_NK}[cc][cc]{$a_{NK}$}
\psfrag{y_1}[cc][cc]{$y_{1}$}
\psfrag{y_N}[cc][cc]{$y_{N}$}
\psfrag{vy_1}[cc][cc]{$v_{y1}$}
\psfrag{vy_N}[cc][cc]{$v_{yN}$}
\psfrag{ms_1}[cc][cc]{$m_{s1}$}
\psfrag{vs_1}[cc][cc]{$v_{s1}$}
\psfrag{ms_K}[cc][cc]{$m_{sK}$}
\psfrag{vs_K}[cc][cc]{$v_{sK}$}
\includegraphics[height=6cm,width=7cm]{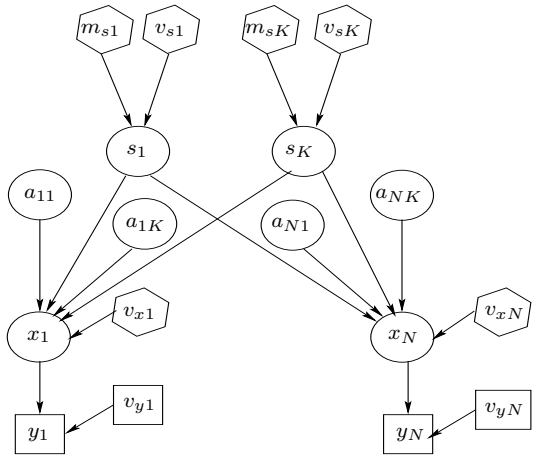}
\caption{Graphical model of the proposed latent variable architecture (i.e. the data model). Nodes
represent random variables in the data model, and arrows indicate
causal relationships between variables. Out of the nodes, rectangles
refer to observed variables: measured fluxes, $y$, and known
measurement errors, $v_{y}$. Circles and hexagons refer to latent
(hidden) variables: clean flux, $x$, along with their variance
parameters, $v_{x}$, mixing coefficients, $a$, and components, $s$,
along with their component-wise means, $m_{s}$, and variance
parameters, $v_{s}$). Hexagons indicate the latent variables on the
top of this hierarchy, for which further hierarchical priors may be
considered.  }
\label{gr_model}
\end{figure}

\subsubsection{The distributional forms employed in the data model}
\label{priors}

In order to be able to infer the parameters of the factor model
(\ref{model}), the form of the distributions of the involved variables,
which may have a key impact on the
subsequent interpretability of the resulting probabilistic model
\citep{mackay03},
need to be defined. That such assumptions are made explicit is indeed
an important feature of the Bayesian framework adopted here.  An
important advantage of this approach is that of the several
alternative forms that could be assumed for these distributional forms
(we could call these ``hypotheses''), the one best supported by the
available data can be determined automatically from the data evidence.
More details can be found below, in \S\ref{me}.

Before proceeding, we note that such assumptions are present in all
automated analysis techniques, although in the case of
non-probabilistic methods they are implicit and fixed. In particular,
it has been shown that the widely used technique of Principal Component
Analysis (PCA) implicitly assumes a standard Gaussian representation
and isotropic Gaussian noise \citep{tb99}.
Although PCA has been useful in a number of ways in astrophysical data
analysis
\citep{connolly95,folkes96,ronen99,madg03a,madg03b}, 
there is no guarantee that the assumptions it makes are appropriate in
our case. In particular, the assumption of a Gaussian distribution for
the spectral elements in our present problem would imply that they are
equally likely to have positive and negative flux values.  This may
make the physical interpretation of the spectra of the individual
stellar populations (factors) problematic. Indeed, the physical
validity of the isotropic structure of the Gaussian noise imposed by
PCA has been questioned by \citet{madg03b}.  

It is therefore worthwhile to consider alternative factor models to
relax such unrealistic constraints, and to introduce more justifiable
constraints (such as the positivity of the factors). This is what we
do in this section.

In this study we consider four factor models with differing
assumptions made regarding the distributional forms employed. These
are summarised in Table~\ref{tab:priors}.  The first three of these
refer to factor models existing in the literature, and the fourth one
has been recently developed by us \citep{harva05}.  All factor models
have been implemented in a Bayesian formulation (the so-called
Variational Bayesian framework, see details in Section \ref{me}) in
order to ensure a consistent formal framework.  The four factor models
are described below.

\textbf{VB-PCA}
The first row of Table~\ref{tab:priors} 
describes the Variational Bayesian
PCA (VB-PCA) data model \citep{bishop99}.  Due to the widespread use
of PCA, we will consider it as the baseline hypothesis.  As already
mentioned, it assumes that the spectral factors of the representation
follow a Gaussian distribution and there is no positivity
constraint. This is summarised in the columns $p(s)$ and $f(s)$
respectively of the first row of Table~\ref{tab:priors}. 
Further, it assumes that the
mixing coefficients of the factors (i.e. the elements of the matrix
$\bd{A}$ of equation (2)) follow a standard Gaussian and finally that
the noise term has an isotropic variance structure (i.e. $v_{x_n}=v_x$
is the same for all $n$). 

\begin{table}
\label{tab:priors}
\begin{threeparttable}
\small
\caption{Distributional assumptions for the various data models}
\begin{tabular}{|c|c|c|c|c|} \hline
Model   & $p(s)$                 & $p(a)$         & $f(s)$         & $v_{x_n}$ \\  
\hline
VB-PCA  & $\cc(s|m_s,e^{-v_s})$  & $\cc(a|0,1)$   & $s$            & $v_x$     \\
VB-FA   & $\cc(s|m_s,e^{-v_s})$  & $\cc(a|0,1)$   & $s$            & $v_{x_n}$ \\
VB-PFA  & $\cc^R(s|0,e^{-v_s})$  & $\cc^R(a|0,1)$ & $s$            & $v_{x_n}$ \\
VB-RFA  & $\cc(s|m_s,e^{-v_s})$  & $\cc^R(a|0,1)$ & $\max(0,s)$    & $v_{x_n}$ \\
\hline
\end{tabular}
\begin{tablenotes}
\item[]{\bf Data Models:} 
VB-PCA: Variational Bayesian PCA; VB-FA: Variational Bayesian Factor
Analysis; VB-PFA: Variational Bayesian Positive Factor Analysis;
VB-RFA: Variational Bayesian Rectified Factor Analysis. The columns
summarise the assumptions for the distributional form of each
variable of Eq.~(\ref{model})
for each of these four data models. Indices
have been dropped. The columns
$p(s)$ and $p(a)$
indicate the form of the probability
density assumed for the elements of $\bd{s}_t$
and $\bd{A}$ respectively.
The column $f(s)$
specifies the form of the function $f()$ of 
Eq.~(\ref{model}), and
$v_{x_n}$ provides the structural form of the noise variance
for the various data models (distinct variances for all data
dimensions $n$ except in the case of VB-PCA).
\end{tablenotes}
\end{threeparttable}
\end{table}

\textbf{VB-FA}
In Table \ref{tab:priors}, the Variational Bayesian Factor Analysis
(VB-FA) data model \citep{bishop99} relaxes the constraint of the
isotropic structure of the Gaussian noise imposed by PCA.  Indeed,
such a constraint may be physically unjustified \citep{madg03b}. All
other assumptions made are in turn identical with those of VB-PCA.

\textbf{VB-PFA}
The above two models do not require the individual factors, which are
spectra of the stellar sub-populations, to be non-negative, which they
need to be in order to be physically acceptable.  The remaining two
models considered here do so.  The third row of Table \ref{tab:priors}
describes the Variational Bayesian Positive Factor Analysis (VB-PFA),
a data model developed by \citet{miskin00}, that, in contrast to
VB-PCA and VB-FA, imposes the positivity constraint. It does so by
assuming that each individual spectrum $p(s)$ (factor) of the
representation follows a zero-mode rectified Gaussian distribution
(essentially a Gaussian restricted to its positive half and
renormalised, defined in Appendix A4).  VB-PFA also restricts the
mixing coefficients $p(a)$ to be positive, assumed to be standard
rectified Gaussians, and maintains the same structural form for the
noise variance $v_{x_n}$ as VB-FA.

The positivity constraint just introduced is meant to facilitate the
interpretability of the factors as component spectra.  Let us observe,
however, that in VB-PFA, the $0-$mode of $p(s)$ would mean that the
recovered components would have a low flux close to zero occupying
most of the spectrum. Although there has been a great deal of interest
in factor models that have such sparsely distributed factors for a
variety of other purposes \citep{mm01,ls01,hg97,socci97,fh99},
including image and text related applications as well as
neurobiological modelling, such a sparsity assumption may be
restrictive and often unjustified for the modelling of stellar
population spectra. However, as shown in \citet{harva05}, modifying
VB-PFA to relax the $0-$mode assumption is technically impossible.
Thus we developed a different way of imposing the positivity
constraint, which also permits the mode of $p(s)$ to be estimated from
the data in a flexible manner. 
We did not discard the hypothesis of
VB-PFA either, since it has been applied to stellar population spectra
in the literature \citep{miskin00} and we do not wish to discard any
existing hypotheses based on possibly subjective intuitions.

\textbf{VB-RFA}
The last row of Table \ref{tab:priors} describes the Variational
Bayesian Rectified Factor Analysis (VB-RFA) data model
\citep{harva05}, where the positivity constraint over the factors is
ensured through the non-differentiable function $f(s)=\max(0,s)$.
With this solution, the factors are now $f(s)$ and they are guaranteed
to be positive. Then the variables $s$ become some dummy-arguments of
$f()$, which are now safely modelled as Gaussians (column $p(s)$ of
4-th row of Table \ref{tab:priors}). Therefore, both the means and
variances of these Gaussians can now be estimated from the data. The
resulting model allows more flexibility regarding the shape of the
factors compared with VB-PFA, while ensuring the positivity of the
factors.  Finally, the mixing coefficients $p(a)$ are assumed to be
distributed as standard rectified Gaussians. Note that in this context
the zero-mode constraint is appropriate since the sparsity of the
mixing matrix $\bd{A}$ translates to assuming a smooth basis
transformation.  The technical novelty and computational advantages of
the VB-RFA approach are detailed in \citet{harva05}.

As it should now be clear, our scope is neither to apply any
off-the-shelf data analysis method in an unjustified manner, nor to
rely on subjective intuitions, as both of these approaches may easily
lead to biased results. Instead, we have constructed four different
hypotheses which we consider equally likely {\it a priori}. The
Bayesian framework allows us to decide which hypothesis is best
supported by the data.

\subsection{The overall model architecture}
The architecture for any of the four data models defined above can be
graphically represented as in Fig.~\ref{gr_model}. This is the joint
density of the observational data and all other variables of
interest. Nodes denote random variables, rectangles refer to observed
variables (i.e. derived from observational data), and circles and
hexagons refer to unobservable variables that we are hoping to infer
from the data through the formulated probabilistic model. Hexagons
denote hidden variables on the top of the hierarchy\footnote{In the
BayesBlocks based implementation
\citep{valpola03} adopted here, a maximal flexibility is ensured
by further formulating hierarchical priors on all variables $m_{s_k}$,
$v_{s_k}$ and $v_{x_n}$, as follows: $p(v_{x_n}) \sim \cc(m_{v_x},
e^{-v_{v_x}})$, with parameters common to all components of the
variable, and these parameters $m_{v_x}$ and $v_{v_x}$ will then have
vague priors in the form of a zero mean Gaussian having a large value
for variance. The definition is analogous for $m_{s_k}$ and
$v_{s_k}$. These additional variables have been omitted from the
diagram on Fig.~\ref{gr_model} for simplicity, as they are part of the
relatively standard technicalities concerning a full Bayesian approach
\citep{valpola05} rather than an essential part of the
application-specific model design.}. Arrows signify generative causal
dependency relations.

\begin{figure}
\begin{center}
\includegraphics[width=0.45\textwidth]{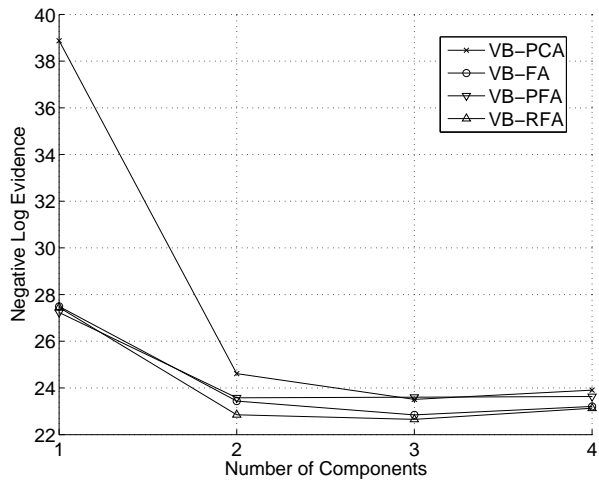}
\caption{The negative log evidence for 
the variational Bayesian analyses of the observed spectra,
as a function of the number of factors (components), \ie\ stellar
sub-populations.  The performance of VB-PCA is
clearly inferior.  VB-RFA has the highest log evidence, with no
significant improvement in increasing $K$, 
the number of components, from two to three.  This shows that the spectra
of early-type galaxies used here can be safely modelled as the sum
of the spectra of two distinct major stellar components, and the inclusion of 
more components is not required by the data.}
\label{costs}
\end{center}
\end{figure}

We now outline the  variational Bayesian estimation
procedure associated with this architecture.

\subsection{Estimation of a data model} \label{me}
Each of the described four data models is estimated using the
variational Bayesian methodology described in this section. The
specific expressions obtained for the VB-RFA data model are provided
in the Appendix.  To make the notation concise, we will refer to the
latent variables by $\bd{\theta}$ and to the data by $\bd{Y}$.  Since
handling the posterior distribution $p(\bd{\theta}\mid \bd{Y})$ is
intractable, we resort to a variational scheme
\citep{jordan99,lap99,valpola05,attias00},
where an approximative distribution
$q(\bd{\theta})$ is fitted to the true posterior. This
is done by constructing a
lower bound of the log evidence, based on Jensen's inequality,
\begin{multline}
\log p(\bd{Y}) 
 = \log \int d\bd{\theta} \, p(\bd{Y}, \bd{\theta})
 = \log \int d\bd{\theta} \, q(\bd{\theta}) 
      \frac{p(\bd{Y}, \bd{\theta})}{q(\bd{\theta})} \\
 \geq \int d\bd{\theta} \, q(\bd{\theta}) 
        \log \frac{p(\bd{Y},\bd{\theta})}{q(\bd{\theta})}
 = \langle\log p(\bd{Y}, \bd{\theta})\rangle_{q(\bd{\theta})}
     - \langle\log q(\bd{\theta})\rangle_{q(\bd{\theta})}
\label{evid}
\end{multline}
where $\langle . \rangle_q$ denotes expectation w.r.t. $q$. 

This inequality holds for any distribution $q(.)$ due to the convexity
of the logarithm. A fully-factorial approximation of the posterior
\citep{mackay03,lap99,attias00,bishop99,bsw02,valpola05}
will be employed here, which is also known to lead to estimation
algorithms that can be implemented efficiently by using local
computations only \citep{lap99,valpola05,valpola03,bsw02} The best
fully-factorial approximate posterior (referred to as the variational
posterior) is computed by functional maximisation of (\ref{evid}), which
can be shown \citep{lap99,mackay03,jordan99} to be equivalent to
minimising the Kullback-Leibler divergence \citep{cover91} of the
variational posterior from the true one. The fully-factorial
variational posterior of the data model in Fig.~\ref{gr_model} is simply of
the form
\begin{eqnarray}
q(\bd{\theta})&=&
\prod_{t=1}^T\prod_{n=1}^N q(x_{nt}) 
\times \prod_{n=1}^N q(v_{x_{n}})
\times \prod_{n=1}^N \prod_{k=1}^K q(a_{nk}) \nonumber\\
&\times& \prod_{k=1}^K\prod_{t=1}^T q(s_{kt})
\times \prod_{k=1}^K q(m_{s_k})q(v_{s_k}).
\label{q}
\end{eqnarray}
Therefore, (\ref{evid}) further decomposes into one plus twice as many
terms as there are latent variables in the data model and the
optimisation w.r.t. the individual variables is separately performed.
The variational posteriors thus obtained, along with the statistics
that are required for an implementation\footnote{All algorithms
investigated here have been implemented as part of the BayesBlocks
software \citep{valpola03, valpola05} and may be made freely available
upon publication.}, are given in Appendix \ref{updates} for the VB-RFA
model. Some of these posterior computations are non-trivial. Readers
interested in more detailed derivations are referred to
\citet{harva05}.

The model estimation algorithm then consists of iteratively updating
each parameter's posterior statistics in turn, while keeping all other
posterior statistics fixed, until a convergence criterion is
achieved. For each update, the evidence bound (\ref{evid}) is
guaranteed not to decrease \citep{jordan99,mackay03}, and convergence to
a local optimum is guaranteed. It can also be shown that due to the
fully-factorial posterior approximation (\ref{q}), all updates are
local, i.e. requiring posterior statistics of the so called Markov
blanket only. That is, for updating any of the variable nodes, the
posterior statistics of only its children, parents and co-parents are
needed. This has been exploited in the efficient implementation
realised in \citet{valpola03}, \citet{bsw02} and \citet{bishop05}, and
has also been adopted in this work. The scaling of the resulting
variational Bayesian estimation algorithm is thus multilinear in the number 
of factors, observations and wavelength bins of the model. These can also be easily followed from
the further details given in
Appendix~\ref{updates}.

The evidence bound (\ref{evid}) can be used both for monitoring the
convergence of the algorithm and, more importantly, for comparison of
different solutions or models. The quantities required for computing
this are given in Appendix
\ref{evidence_eq}.

\subsection{Using a data model 
to recover missing elements of the data}\label{missdat2} 

We have
already stressed the advantage of the probabilistic Bayesian framework
adopted here in terms of the ability of the resulting data models to
predict and recover incomplete data.  Recovering missing values can be
achieved by computing the posterior of the missing entries given the
observed entries:
\begin{equation}
q(\bd{y}_t^m | \bd{y}_t^o)=\int d\bd{\theta} 
              q(\bd{\theta})p(\bd{y}_t^m|\bd{\theta}).
\label{missing}
\end{equation}
An approximate mean value of this posterior provides the imputation
for the missing entry.  This is discussed in Appendix~\ref{miss}.

\subsection{Using a data model for recovering missing 
elements from new measurements}\label{missdat3} 

Due to the fully generative nature of the adopted framework, once the
data model parameters are estimated from a training data set,
inferences and predictions can be made for new, previously unseen
measurements. This involves performing the inference for those terms
of the variational posterior $q(\bd{\theta})$ that are conditional on
the type of new measurements (new galaxy spectra in the same range or
various sub-ranges of wavelength for the same galaxies).  Using these,
as well as the remainder of terms of $q(\bd{\theta})$ as estimated
from the training set, we can then proceed to computing
(\ref{missing}). Results will be presented in Section 4.3.

\section{Results}

One of the strengths of the Bayesian evidence framework employed in
this study is that the selection of the best-supported hypothesis can
be automatically achieved, 
in a quantitative manner, based on the data
evidence. This is known as model selection.

\subsection{Model selection results: how many major 
stellar populations do we need?}  

Here we seek a criterion to examine which data model performs the
best, as well as to find the dimensionality of the representation
space. We would like to have a compact model (small number of factors)
both to avoid over-fitting the data, and also for the results to be
physically interpretable.  At the same time, we would like to keep an
adequate number of factors to guarantee a good approximation of the
data. Apart from the interpretability issue, the Bayesian evidence
framework can be used to automatically select the optimal model and
its order from those investigated.  Our preferences, motivated by the
purpose of the application (e.g. data interpretation, missing value
prediction), expressed as prior belief, would then be used to weight
the evidence obtained from the data, in order to make a final Bayesian
decision on the order of the model.

Fig.~\ref{costs} shows the negative of the log evidence for the four
data models (strictly, the negative log evidence bound), versus the
model order, $K$ (i.e. the number of components), for the observed
spectra (Fig.~\ref{spec}). 
It can be seen that the VB-RFA model outperforms the other
three models. We also see that increasing the number of factors beyond
two does not significantly increase the evidence, which suggests that
the bulk of spectra of an early-type galaxy can be reasonably well and
compactly described by two factors or components, each interpretable
as a stellar sub-population. This is useful for input into
more detailed physical models of early-type galaxy spectra. We will
also see in \S\ref{physical-interp}
 that the $K=2$ models produce two factors that
are both physically meaningful, while increasing the model order to
$K=3$ does not bring any further interpretable factors.

\begin{figure*}
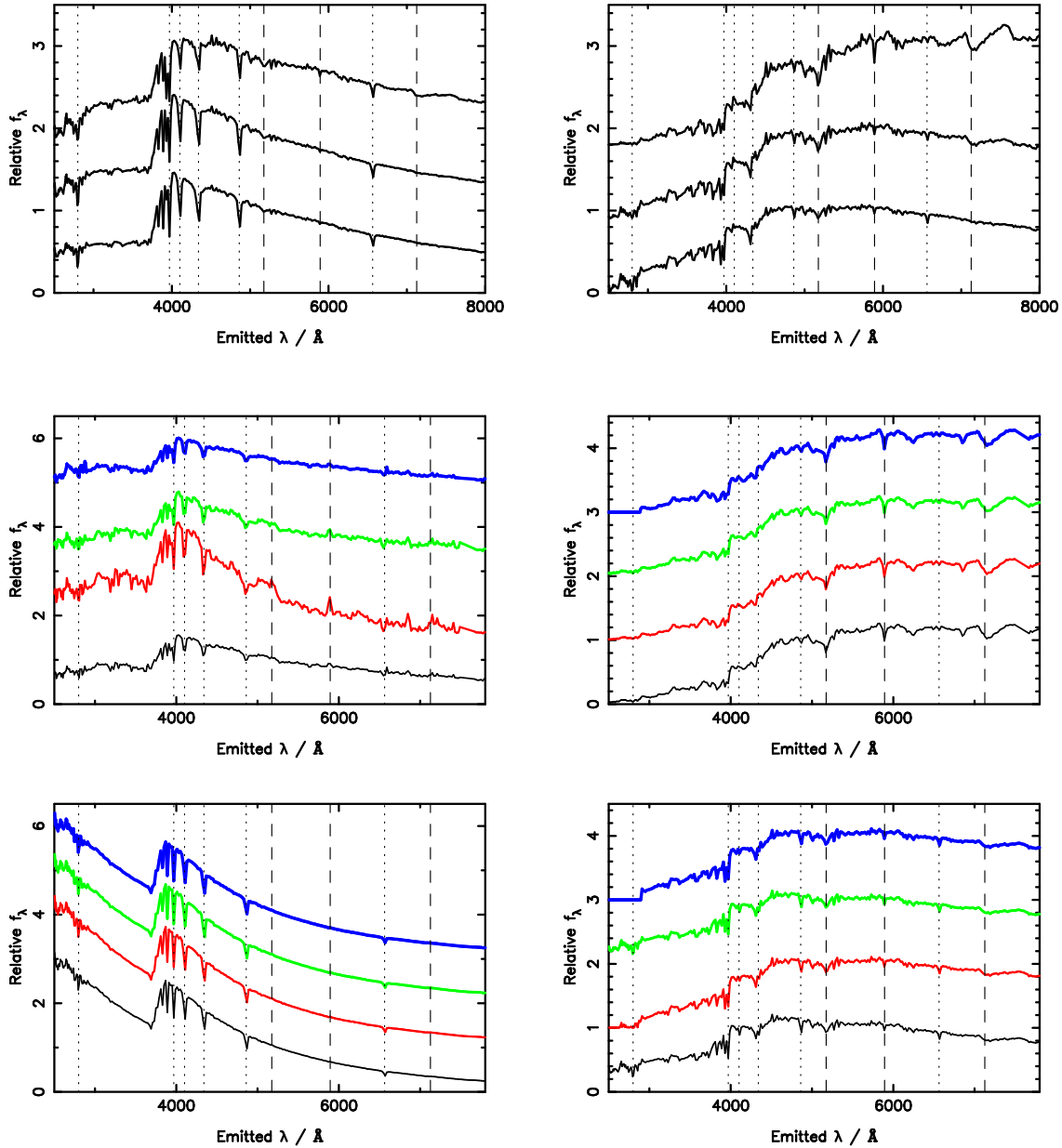

\begin{center}
\includegraphics[angle=-90,width=0.85\hsize]{fig4a.eps}
\includegraphics[angle=-90,bbllx=320,bblly=28,bburx=583,
bbury=746,clip=,width=0.85\hsize]{fig4b.eps}
\caption{{\bf Analysis of the two-component data model.} 
{\it Top:} 
Synthetic stellar population spectra \citep{jimenez04}: {\it
Top Left:} Spectra of a population of age 0.7~Gyr, where metallicity, 
from bottom to top, is $Z=$0.2, 1.0 and 2.5 \Zsolar; {\it
Top Right:} age = 10~Gyr, same metallicities. The dotted lines mark
some of the absorption features in the spectrum which are typically
strong in young stellar populations, and the dashed lines mark some of
the absorption features which are typically strong in old, metal-rich
stellar populations. From left to right, the absorption line species
are: MgII (2799 \AA), H$\varepsilon$ (3970 \AA), H$\delta$ (4102
\AA), H$\gamma$ (4340 \AA), H$\beta$ (4861 \AA), Mgb (5175 \AA),
NaD (5893 \AA), H$\alpha$ (6563 \AA), TiO (7126 \AA). {\it Middle:}
The two components found from the various different linear independent
basis transformation analyses, with the number of components $K=2$, of
the observed early-type galaxy spectra. The methods are, from bottom
to top: VB-FA, VB-PFA, VB-PCA, VB-RFA. The recovered spectra are
convincingly disentangled into one component (middle left) with young
stellar population features (MgII, H$\varepsilon$, H$\delta$,
H$\gamma$, H$\beta$, H$\alpha$: dotted lines) and shape, and a second
(middle right) with the features (Mgb, NaD, TiO: dashed lines) and
shape of an old, high-metallicity stellar population.  {\it Bottom:}
The two components found from the same analyses, but from the linear
superposition of two synthetic stellar populations. The various curves
refer to the same parameters as in the middle panels. The UV spectrum
of the `young' component recovered here rises more steeply than for
the same component recovered from the observed spectra, due to the
inclusion, in the models,
of a very young (0.03 Gyr) synthetic stellar population,
which is not seen in our observed spectra.}
\label{modelspecr2}
\end{center}
\end{figure*}

\subsection{Physical interpretation of the results}
\label{physical-interp}

Here, we compare the results of the data-driven method outlined 
so far with an entirely independent determination of
the star-formation history of our 21 early-type galaxies. 

In this more conventional approach,
we have fitted two-component detailed physical 
models, constructed from the single stellar population models of
\citet{jimenez04}, to the observed long-baseline spectra of our
sample, as shown in Fig.~\ref{spec}.  
The ages, metallicities and relative mass fractions of the
component stellar populations are allowed as free parameters. We
construct the two-component spectra as the linear superposition of the
two component spectra (\ref{modflux}), and allow ages 0.01-14.0 Gyr,
and metallicities $Z=$ 0.01, 0.2, 0.5, 1.0, 1.5, 2.5 and 5.0 times the
solar value (\Zsolar). Again, the relative mass fractions have to be
jointly constrained, $m_i + m_j = 1$, but $m_{i}$ is allowed to vary
from 0 to 1, in steps of 0.02. A minimum-\xs\ fit was used to
determine the best-fit values of $m_{i}, Z_{i}, t_{i}, m_{j}, Z_{j}$
and $t_{j}$. The whole parameter space was searched, with the
best-fitting parameter values corresponding to the point on the
age$_{i}$-age$_{j}$-Z$_{i}$-Z$_{j}$-$m_{i}$-$m_{j}$ grid with the
minimum calculated {\xs}. Details of the fitting process are given in
\citet{nolan02} and \citet{nolan03}.

\subsubsection{Analysis of the observed spectra 
with the two-component data model}\label{r2}

In Fig.~\ref{modelspecr2}, synthetic spectra \citep{jimenez04} of
young (0.7 Gyr, left) and old (10 Gyr, right) stellar populations are
shown, for metallicities 0.2, 1.0 and 2.5 \Zsolar, in
each of the upper panels, from top to bottom respectively. Below these
are shown the spectra of the individual components, recovered
from the four analysis methods described above, for both the observed
galaxy spectra (middle panels), and linear superpositions of pairs of
synthetic spectra (bottom panels).

\begin{figure}
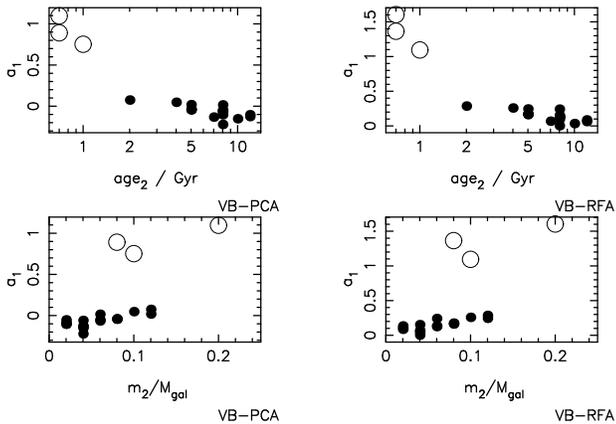

\begin{center}
\includegraphics[width=2.75cm,angle=-90]{fig5a.epsf}
\includegraphics[width=2.75cm,angle=-90]{fig5b.epsf}
\caption{Scatter-plots showing the correlation of {\it (top)}
the age of the younger stellar population and {\it (bottom)} the mass
fraction of the smaller stellar population determined from
two-component fitting to the observed spectra with the weight of the
first component ($a_1$) of the VB-PCA (left) and VB-RFA (right) linear
basis transformation analyses of the observed spectra, with $K= 2$. A
high value of $a_1$ clearly corresponds to a substantial young (\ls\ 1
Gyr) stellar population. The open circles are for those galaxies with
a secondary population of age less than 1~Gyr. These plots also show
that the worst-performing (VB-PCA) and best-performing (VB-RFA)
algorithms in this case yield almost identical results.}
\label{corrplots-r2}
\end{center}
\end{figure}

In general, the similarity between the recovered components, and
the corresponding model spectra, is striking, for all four methods. In
the younger
components displayed on the left hand of Fig.~\ref{corrplots-r2}, 
the 4000~\AA\ break, typical of a mature (\gs 1
Gyr), is not seen, but strong Balmer absorption line features
(H$\varepsilon$ (3970 \AA), H$\delta$ (4102
\AA), H$\gamma$ (4340 \AA), H$\beta$ (4861 \AA)) can be clearly
identified. These are typical features of the A~stars which dominate
the emission from young ($<$ 1 Gyr) stellar populations. In the second
(older) component, however, the 4000 \AA\ break is apparent, as are
other features typical of mature, metal-rich stellar populations,
e.g. absorption features corresponding to Mgb (5175 \AA), NaD (5893 \AA)
and TiO (7126 \AA). We can rule out either component being representative predominantly of either  
AGN emission or dust absorption as the characteristic shapes and features of these spectra would 
be very different from those we recover.

However, the VB-PCA algorithm seems to ``find''  a spurious
sub-component with younger features (middle left panel) that has some
features from old, metal-rich stellar populations (notably NaD).

\begin{figure}
\centering
\includegraphics[width=4.0cm,angle=-90]{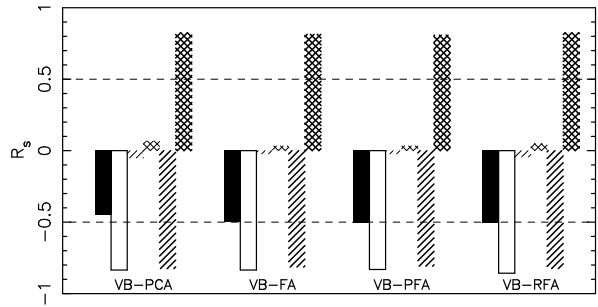}
\caption{Spearman's rank correlation 
coefficient (R$_{S}$) for the first component ($a_1$) of the four linear
basis transformation analyses with K $=$ 2, of the observed spectra
and the best-fitting parameters of the two-component synthetic stellar
population fits to the data: solid: age of the older population;
outline: age of the younger population; light hatched: metallicity of
the older population; light cross-hatched: metallicity of the younger
population; dark hatched: mass fraction of the smaller population;
dark cross-hatched: mass fraction of the dominant population. $\bmod$ R$_{S} >
0.5$ indicates a strong correlation.  }\label{tabrs2}
\end{figure}

In Fig.~\ref{corrplots-r2}, we show results for VB-RFA method (right
panels), which has the the highest log evidence values, together with
the VB-PCA method (left panels), which has the lowest, for
comparison. We plot the correlation between the first (`younger')
component ($a_1$) and the age of the younger stellar population
(age$_2$, top panels) and the mass fraction of the secondary stellar
population component ($m_2/M_{\rm gal}$), as determined from the
two-component model-fitting described above (bottom panels). The open
points are those galaxies which contain a significantly young stellar
population (\ls 1~Gyr).

Fig.~\ref{tabrs2} plots the results of the Spearman's rank
correlation test for the weight of the first component $a_1$ and the
age, metallicity and mass fraction of the two fitted stellar
populations for the observed spectra with $K \!=\! 2$. A value of the
Spearman's rank correlation coefficient $|R_S|\!>\! 0.5$ indicates a
strong correlation between the parameters. It can be seen that there
is a strong correlation between $a_1$ and the ages of the younger
stellar populations, and also between $a_1$ and the stellar mass
fractions ($m_1M_{\rm gal}$, $m_2/M_{\rm gal}$) of the component
stellar populations. The correlation between $a_1$ and age$_1$ is
weak, and there is no correlation between $a_1$ and the metallicities
of the component stellar populations. The trends of the correlations
with $a_2$ follow similar, but opposite, trends to those of
$a_1$. This is not generally true for factor analysis, but arises here
as a consequence of the nature of our data set.

\begin{figure*}
\begin{center}
\includegraphics[width=12.0cm,angle=-90]{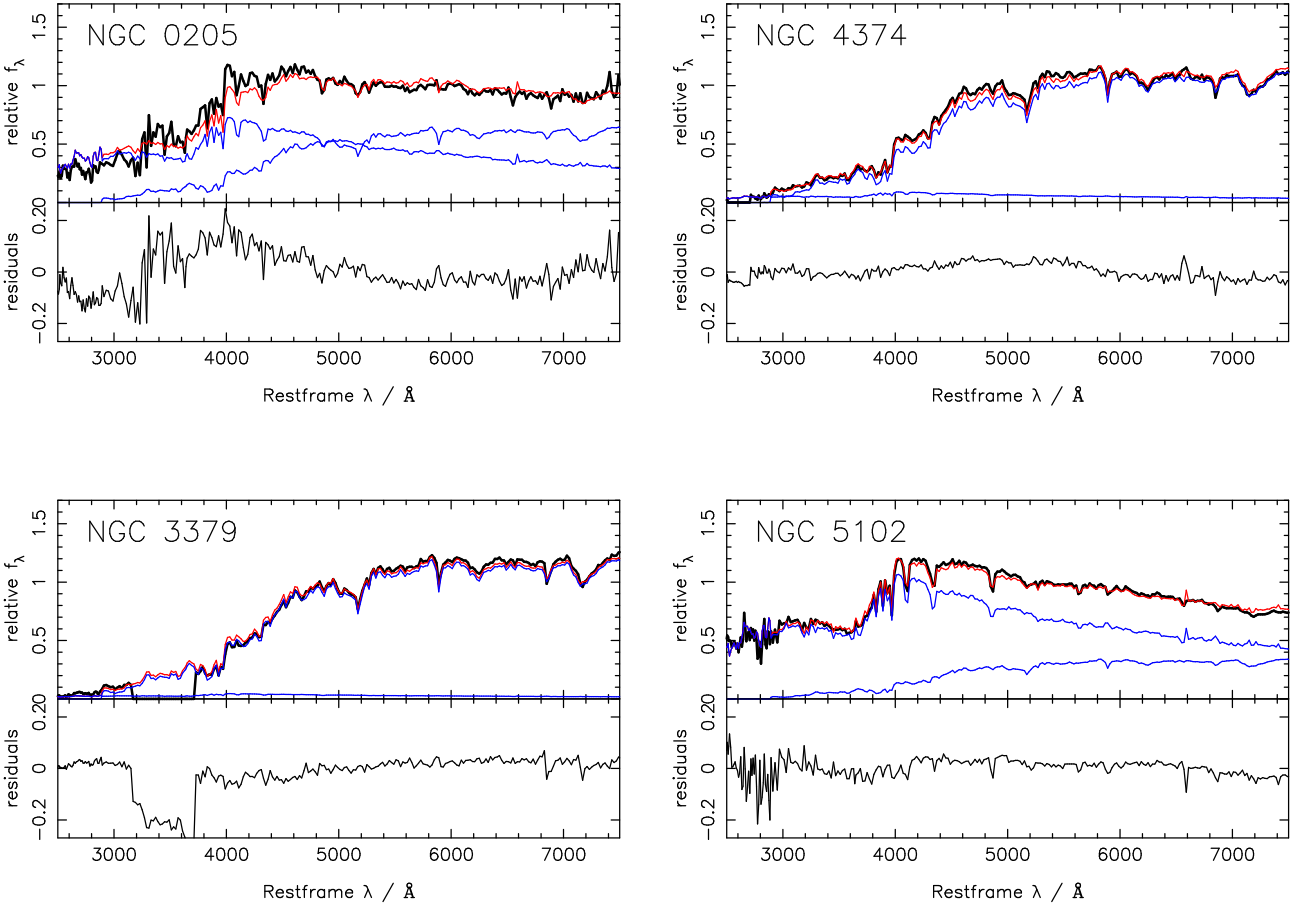}
\caption{Rest-frame spectra of four of the 21 galaxies 
in our sample (thick black line), with the reconstructed data model
and the two constituent data model components (thin lines)
superimposed. The data model convincingly reproduces the observed
spectra.}\label{recon}
\end{center}
\end{figure*}

The rest-frame spectra of four of the galaxies in our sample, together
with the reconstructed data model and the two constituent data model
components are presented in Fig.~\ref{recon}. The data model
convincingly reproduces the observed spectra.

\subsubsection{Analysis of the synthetic spectra with the 
two-component data model}

\begin{figure}
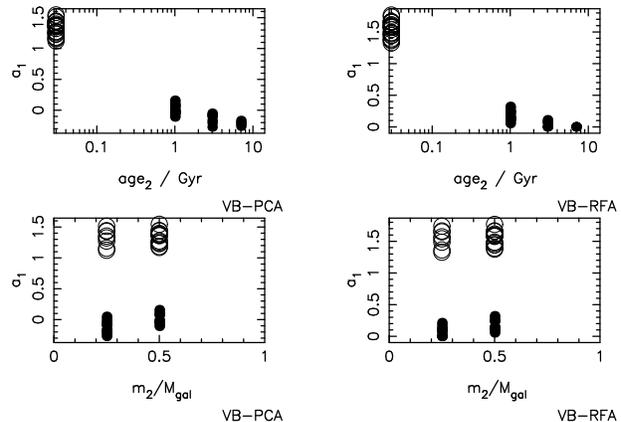

\begin{center}
\includegraphics[width=2.75cm,angle=-90]{fig8a.epsf}
\includegraphics[width=2.75cm,angle=-90]{fig8b.epsf}

\caption{Scatter-plots showing the correlation of {\it (top)}
the age of the younger stellar population and {\it (bottom)} the mass
fraction of the smaller stellar population determined from the
synthetic spectra fitting with the weight of the first component of
the VB-PCA and VB-RFA linear basis transformation analyses methods
($a_1$), and $K= 2$. The open circles are for those galaxies with a
secondary population of age less than 1 Gyr. A high value of $a_1$
clearly corresponds to a significant young ($\ll 1$ Gyr) stellar
population.}
\label{corrplots-s2}
\end{center}
\end{figure}

\begin{figure}
\centering
\includegraphics[width=4.0cm,angle=-90]{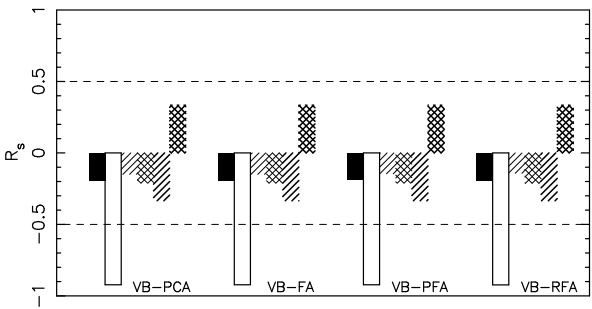}
\caption{Spearman's rank correlation coefficient (R$_{S}$) for the first
component ($a_1$) of the various linear basis transformation analyses
with $K = 2$, of the two-component synthetic spectra and the
best-fitting parameters of the two-component synthetic stellar
population fits to the data: solid: age of the older population;
outline: age of the younger population; light hatched: metallicity of
the older population; light cross-hatched: metallicity of the younger
population; dark hatched: mass fraction of the smaller population;
dark cross-hatched: mass fraction of the dominant population. $\bmod$ R$_{S} >
0.5$ indicates a strong correlation.}\label{tabss2}
\end{figure}

Here we present the results of the analysis of the two-component
synthetic spectra, constructed as described in \S\ref{synthetic}. They
represent every possible combination of the assigned ages,
metallicities and relative stellar mass fractions, and thus allow us
to investigate a wider range of known age and metallicity combinations
than the observed spectra. It should be noted that not all
combinations (for example, those with two components $<$ 1 Gyr old)
represent what we might expect to find in the total population of real 
early-type galaxies, although there could be local regions within these 
galaxies where, for example, young stars dominate the stellar population. 
We wanted to cover the complete range of
possibilities in order to avoid making any assumptions about the
stellar populations of early-types, which, as revealed even in our
small sample of twenty-one spectra, can be quite varied. 

In this experiment, no noise has been added to the synthetic spectra. This is because the 
noise from variations in the spectral shape caused by differences in the 
metallicity is higher than any noise contributions which represent measurement errors.

The bottom panels of Fig.~\ref{modelspecr2} show the spectra of the
components recovered using the same four analysis methods as used for
the observed spectra. Again, the first component has a shape
and features that are similar to those
of a young stellar population, and the second component,
 those of a mature population. The shape of the UV part of the
spectrum, together with the featureless nature of the spectrum above
$\sim$5000 \AA, are typical of very young stellar populations ($\sim$
a few x 10 Myr). These populations are present in the synthetic data
set, but not in significant quantities in the observed data set, hence
the differences between the recovered data components.

In Fig.~\ref{corrplots-s2}, as in Fig.~\ref{corrplots-r2}, we see
that a high value of $a_1$ (the weight of the `younger' component from
the Bayesian data model) corresponds to the presence of a $<$1~Gyr
stellar population. Fig.~\ref{tabss2} plots the Spearman's rank
correlation test results for $a_1$ and the age, metallicity and mass
fraction of the synthetic populations. Here, there is an even stronger
correlation between $a_1$ and the age of the younger population than
in the case of the observed spectra, and also a strong correlation
between $a_1$ and the age of the older population, but no significant
correlation between $a_1$ and the mass fraction of the secondary
population. However, in this exercise with synthetic spectra, we
allowed only two values for $m_2/M_{\rm gal}$ (0.25 and 0.5), which
does not really allow for a meaningful determination of the
correlation coefficient.

These results confirm that the value of $a_1$ is a clear indicator of
the presence of a significant young stellar population, and is
particularly unambiguous in discovering components with very young
($<$1 Gyr) stellar populations, as we found in \S\ref{r2}.

\subsubsection{Analysis of the three-component data model}

\begin{figure*}
\begin{center}
\includegraphics[angle=-90,width=0.9\hsize]{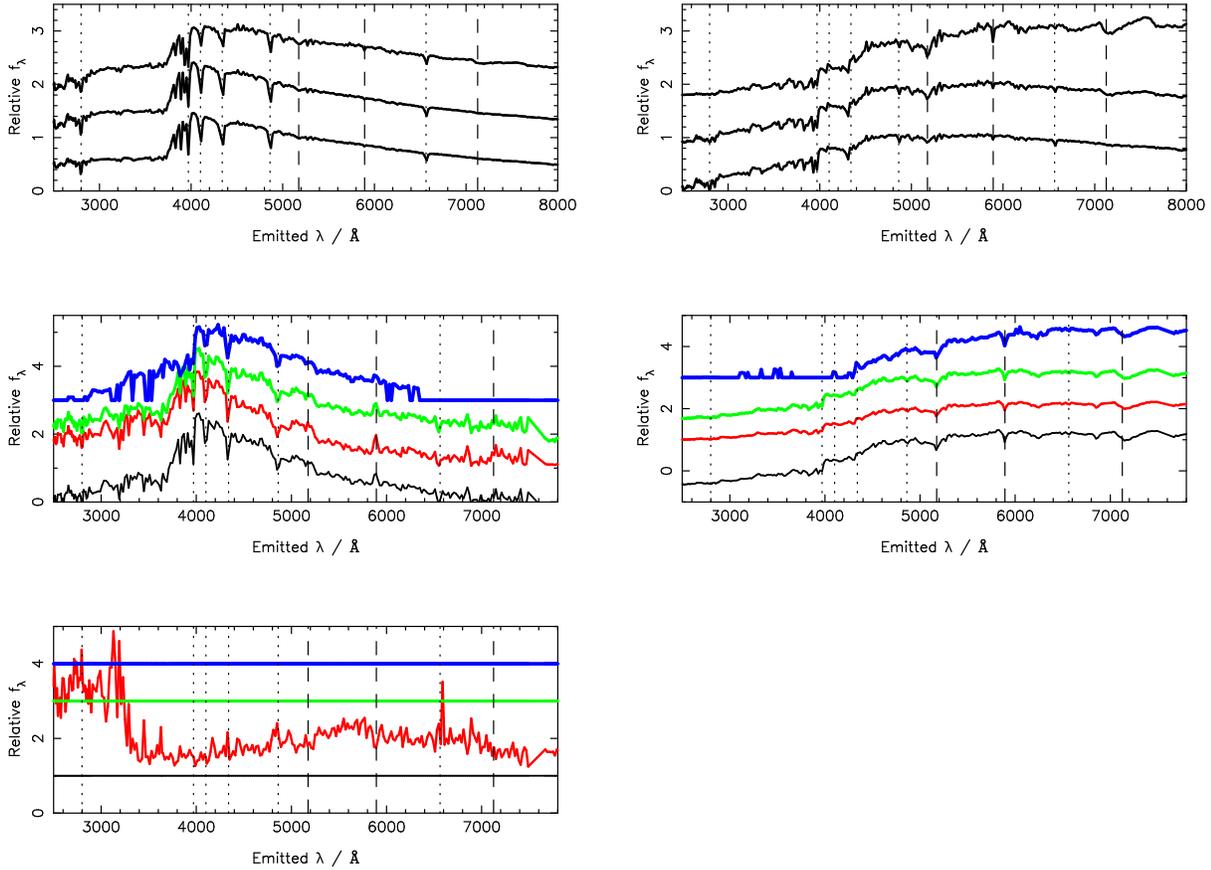}
\caption{{\bf Analysis of the three-component data model.}
{\it Top panel:} 
Synthetic stellar population spectra \citep{jimenez04}: {\it top left:}
for a stellar population of age 0.7~Gyr, with from bottom to top,
metallicity $Z=$ 0.2, 1.0 and 2.5 \Zsolar; {\it top right:} age =
10~Gyr, same metallicities. The dotted lines mark some of the
absorption features in the spectrum which are typically strong in
young stellar populations, and the dashed lines mark some of the
absorption features which are typically strong in old, metal-rich
stellar populations. From left to right, the absorption line species
are: MgII (2799 \AA), H$\varepsilon$ (3970 \AA), H$\delta$ (4102
\AA), H$\gamma$ (4340 \AA), H$\beta$ (4861 \AA), Mgb (5175 \AA),
NaD (5893 \AA), H$\alpha$ (6563 \AA), TiO (7126 \AA). {\it Middle
and bottom panels:}
The 3 components found from the various different linear independent
basis transformation analyses of the model stellar populations, with
$K=3$. The methods are, from bottom to top: VB-FA,
VB-PFA, VB-PCA, VB-RFA. The first two recovered spectra are broadly
identified with the features of a young and old stellar population in
the same way as the $K= 2$ components. The third component (bottom)
possibly contains `residuals', resulting from the fact that the first two
components are not a complete detailed reconstruction of the true
stellar population components.}
\label{modelspec-s3}
\end{center}
\end{figure*}

Fig.~\ref{costs} shows that increasing the number of component spectra
in the Bayesian models from two to three does not significantly
increase the evidence for any data model. Indeed, for the positive
factor analysis, the evidence decreases as $K$ increases from two to
three. These results suggest that the bulk of the stellar content of
early-type galaxies can be represented by two components, one which
represents contributions from old, metal-rich stellar populations, the
other representing contributions from young populations
(Fig.~\ref{modelspec-s3}). It is not easy to find a physical
interpretation for the third component. Indeed, the third component is
probably a combination of residuals from the sum of the spectra of two
populations, combined with correlated components of observational
errors.

\begin{figure}
\centering
\includegraphics[width=12.0cm,angle=-90]{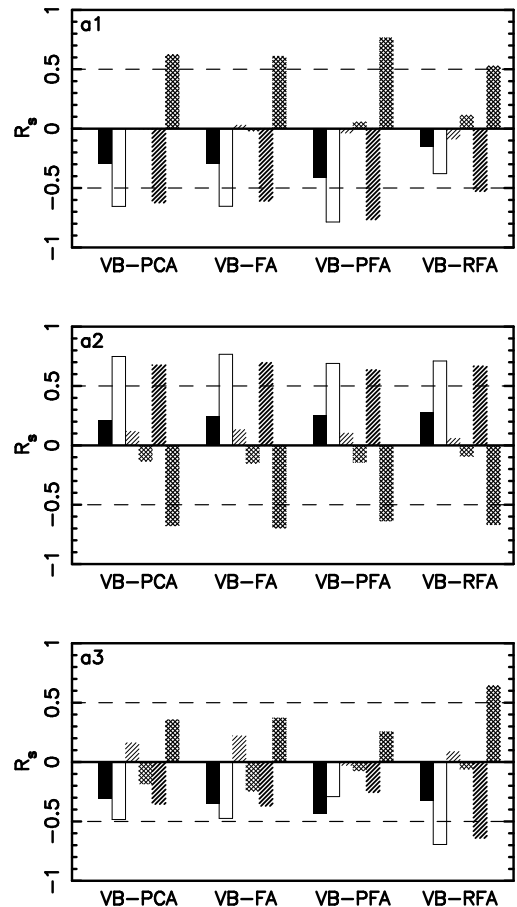}
\caption{Spearman's rank correlation coefficient (R$_{S}$) for the
first component ($a_1$) of the various linear basis transformation
analyses with $K= 3$, of the observed spectra and the best-fitting
parameters of the two-component synthetic stellar population fits to
the data: solid: age of the older population; outline: age of the
younger population; light hatched: metallicity of the older
population; light cross-hatched: metallicity of the younger
population; dark hatched: mass fraction of the smaller population;
dark cross-hatched: mass fraction of the dominant population. $\bmod$ R$_{S} >
0.5$ indicates a strong correlation.}\label{tabrs3}
\end{figure}

The results of the analysis performed on the observed spectra are
presented in Fig.~\ref{tabrs3}. The components $a_1$ and $a_2$ correlate significantly with the age of
the younger population (age$_2$), except for the VB-RFA, where the correlation is with $a_3$ rather than $a_1$. The relative mass fraction of the two stellar population components ($m_1$ and $m_2$) also correlate with $a_1$ and $a_2$, and, for the VB-RFA, with $a_3$ as well.

For the synthetic two-population spectra, age$_2$ correlates strongly
with $a_1$ and $a_2$, for all the analysis methods, and also with
$a_3$ for the VB-RFA. However, there is no strong correlation ($\bmod$
R$_{S} > 0.5$) for any of the data models with the relative mass
fraction of the two populations, although the metallicity of the
younger population ($Z_2$) correlates strongly with $a_3$ in the case
of the VB-PCA and VB-FA methods.

The differences between the correlation coefficients for the observed
and synthetic spectra most likely reflect the different distributions
of age, metallicity and mass fraction in the two spectral samples. For
example, the observed spectra are all dominated by high-metallicity
components (2.5 and 5.0 \Zsolar), with small contributions from very
low-metallicity populations (0.01 \Zsolar). Only one of the galaxies
contains stellar populations with intermediate metallicities. In
contrast, the synthesised spectra can take the whole range of
combinations of the seven allowed metallicities.

\subsubsection{Interpretation of the eigen-components of PCA}

Although we have managed to identify components with PCA here, it is a well known property of PCA that the solution it provides is rotation-invariant. That is, the eigen-components which have equal variances 
can be multiplied by an arbitrary orthogonal matrix to produce another equally 
good solution (in terms of satisfying the same objective function). Therefore 
PCA is not guaranteed to achieve a successful separation of the underlying independent 
source signals that we are looking for. Methods that take into account higher 
order dependencies need to be used for that purpose \citep{ICABook1}.

The only 'lucky' case where PCA can identify components is if the underlying 
components have suitably different variances. This is true for the work presented here, but there is no reason why this would hold for data sets of galaxy spectra in general. Therefore, even if PCA 
has been successful here, it comes with no guarantees.

Another way of looking at this problem is through the probabilistic formalism. 
PCA's implicit assumption (as shown in detail in \citet{tb99}) is that the 
components it will find are Gaussian. Making this assumption explicit, by
the probabilistic formalism that we adopted, highlights the lack of 
plausibility of this assumption: the spectra that we are looking for are
not Gaussian, instead they are positive valued only, and have a rather skewed 
distribution. This renders the use of PCA very problematic in applications 
that require a clear interpretation of the components. 

In turn, by adopting the probabilistic formalism where any assumption in the 
data model must be made explicit, we are able to specify the distribution that 
corresponds to components that we are looking for. Therefore the chances that we 
can find what we are looking for, and not something else, are considerably increased. The 
ultimate conclusion is of course made based on visual inspection of the recovered  
components, made by human experts.

\subsection{The prediction of missing values}
As a more objective criterion for the assessment of our data modelling
compared to the results obtained by the fitting of detailed spectral
models, here we investigate the predictive ability of the modelling
approach when faced with missing elements of the spectra.

\begin{figure}
\centering
\includegraphics[height=6.5cm,width=6.5cm]{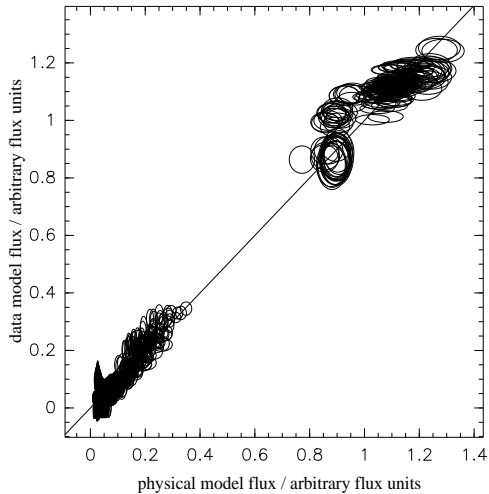}
\caption{Predictions for the missing values provided by the VB-RFA
model versus the physical model. Values of the x-radii of the ellipses
represent errors (one standard deviation) obtained from data
imputation using synthetic physical models.  The y-radius values
correspond to one standard deviation of the posteriors of $\bd{x}$, as
obtained using VB-RFA. There are no missing values in the observed
spectra in the central region of the wavelength range considered here,
which accounts for the gap in the plotted data.}
\label{fig:realmiss}
\end{figure}

We begin with demonstrating the recovered values for the missing
elements along with their predicted uncertainties (posterior
variances), as obtained from VB-RFA on the small real data set,
compared with those from the best-fitting physical model. This is
shown on the scatter plot in Fig.~\ref{fig:realmiss}. The correlation
coefficient between the posterior mean predictions of the data model
and the values imputed from the best-fitting physical model is
0.989, the SNR being 9.3. This is a strong indication that the
data-driven analysis and the physical analysis are well matched.

However, in the case of the real data, the true values of the missing
entries are unknown. Therefore, in the following section, we employ
synthetic data to perform a more thorough and controlled objective
evaluation of the predictive capabilities of our probabilistic data
model model.  Missing entries were simulated, and the true values were
used for evaluation only. The synthetic data set was more diverse than
the real one, as a variety of parameter combinations that can be simulated,
in this process, may not appear in real spectra. Therefore, a wealth of
situations were created on which the generalisation abilities of the
proposed method were tested. A global and fixed noise level of
$\sigma_{nt}^2=0.01$ was used in the data models in these
experiments \footnote{Here, $\sigma_{nt}$ can be used to both control
the compactness of the representation and to simulate a level of
measurement uncertainly similar to that encountered in real data.}. We
studied the prediction performance for both two and three
component models ($K=2$, $K=3$) on new,
previously unseen measurements, in the two scenarios as reported
below. In this setting, the evidence bound peaked at $K = 3$, as can
be seen in Figures
\ref{fig:specinf} and \ref{fig:waveinf}.

\subsubsection{Predictions for previously unseen spectra in the 
same wavelength range}

As the spectra of more and more galaxies become available and are
added to public archives, the ability to make predictions for
previously unseen spectra, that have a small percentage of missing
values within the same overall wavelength range, is an important
practical requirement.

\begin{figure}
\centering
\includegraphics[width=7cm,height=7cm]{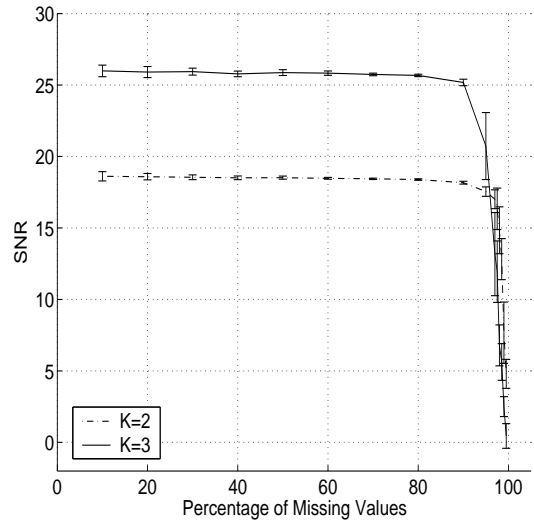}
\caption{Predictive performance from
previously unseen spectra using a VB-RFA model with $K=2$ (bottom) and
$K=3$ (top), plotted against the percentage of missing entries from
the test-set. }
\label{fig:specinf}
\end{figure}

It is therefore desirable to be able
to infer reliable estimates of missing values, based on a few
measurements and the probabilistic representation already created from
previously seen spectra. In this scenario, Fig.~\ref{fig:specinf}
shows the prediction results obtained by VB-RFA when varying the
percentage of missing entries in the test set. The 
predictions are surprisingly accurate up to a high
percentage of missing entries among the previously unseen spectra. 
Effectively, the high redundancy across the spectra, due to the
fact that mature galaxies are very similar to each other, makes this
scenario relatively easy to handle.

\subsubsection{Predictions for previously unseen wavelength bins}

The second scenario investigated here is when the predictions are to
be made for missing elements for previously unseen values
of wavelength, albeit within the same overall wavelength range.
 
This is a more difficult task, since the flux varies significantly
across different wavelength ranges. We conducted experiments in which
the overall wavelength range for each spectrum
was split randomly into training sets (regions without missing
values in our case, though they need not be so) and test sets, and each model
was trained on the training set.
Then, to ensure that we retain the 
missing-at-random \citep{gj94} assumption,
for randomly picked ranges in the
test set, ``missing'' values were predicted 
based on the trained model and the prediction errors
were averaged over all missing values. 
This procedure was repeated 10 times
 with 50 different combinations of
training and test sets for each spectrum.

\begin{figure}
\centering
\includegraphics[width=7cm,height=7cm]{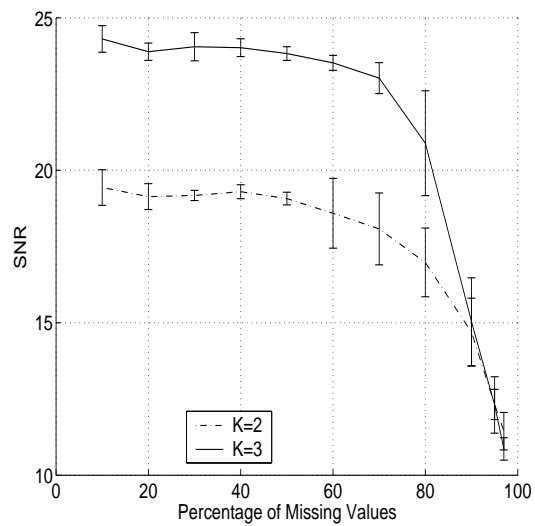}
\caption{Predictive performance from
previously unseen wavelength bins using a VB-RFA model with $K=2$ (bottom)
and $K=3$ (top).  As before, the x-axis indicates the percentage of
missing entries from the test-set. }
\label{fig:waveinf}
\end{figure}

The results are shown in Fig.~\ref{fig:waveinf}. The percentage of
missing values was varied between experiments as is indicated on the x
axis.  With a complete training set, the predictions are still pretty
accurate even when up to 60\% of the bins 
from the test set are missing.

\section{Finding E+A galaxies}

So far we established generic data-driven procedures for finding
evidence of more than one stellar sub-populations, significantly
differing in age and metallicity-related characteristic features, in
the integrated spectra of galaxies. Here we show an application to a
particular data mining project which would benefit from this method, once 
large sets of galaxy spectra are available for public use.

The determination of the star formation history of early-type galaxies has
important implications for the much-debated issue of their formation and
evolution. Observational evidence suggests that at least some elliptical
galaxies already have the bulk of their stars at high redshift ($z\!>\!2$) 
\citep[\eg][]{Ren98,nolan03}, but in hierarchical structure formation
models, ellipticals can form at lower redshift from late-type merging
\citep[\eg][]{WR78,KWG93}. 
A vital link in understanding how early-type
evolution is driven, and under what environmental conditions, is the study
of E$+$A galaxies. 

The spectra of these galaxies are characterised by strong hydrogen Balmer
absorption-lines together with a lack of significant [OII]3272 \AA\
emission, implying a recent (\ls 1 Gyr) starburst, but no significant
on-going star formation. The overall shape of the
spectral energy distribution of these galaxies is that expected for an
early-type galaxy. They may represent an intermediate stage between
disc-dominated rotationally supported systems and spheroid-dominated,
pressure-supported systems.

The physical mechanisms responsible for the triggering and subsequent
truncation of star-formation in E$+$A galaxies are currently
unclear. They may be the result of disk-disk merging (or galaxy
interactions), or cluster-related phenomenon, e.g. ram-pressure
stripping, tidal stripping.  However, although at intermediate
redshift, it has been suggested that E$+$A-type galaxies are prevalent
in clusters \citep[\eg][]{tran03}, which argues for a cluster-related
mechanism, at low-redshift ($z\! \sim\! 0.1$), E$+$As occur in
lower-density environments, which is more indicative of merging or
interaction processes \citep{Zab96,blake04}. Statistical studies of
the environments, luminosities and detailed morphologies of these
galaxies, across a range of redshifts, are necessary in order to
uncover the physical processes governing their star-formation history.

The possible relationship between early-type evolution and E$+$A galaxies
has not as yet been well-studied, as E$+$A systems are rare ($<$1 percent
of the overall zero-redshift galaxy population, \citet{Zab96}). However,
recent and on-going large-scale surveys such as the SDSS and the 2dFGRS offer
the opportunity to study these galaxies in a statistically meaningful way,
if they can be reliably identified.

\begin{figure}
\begin{center}
\includegraphics[width=12.0cm,angle=-90]{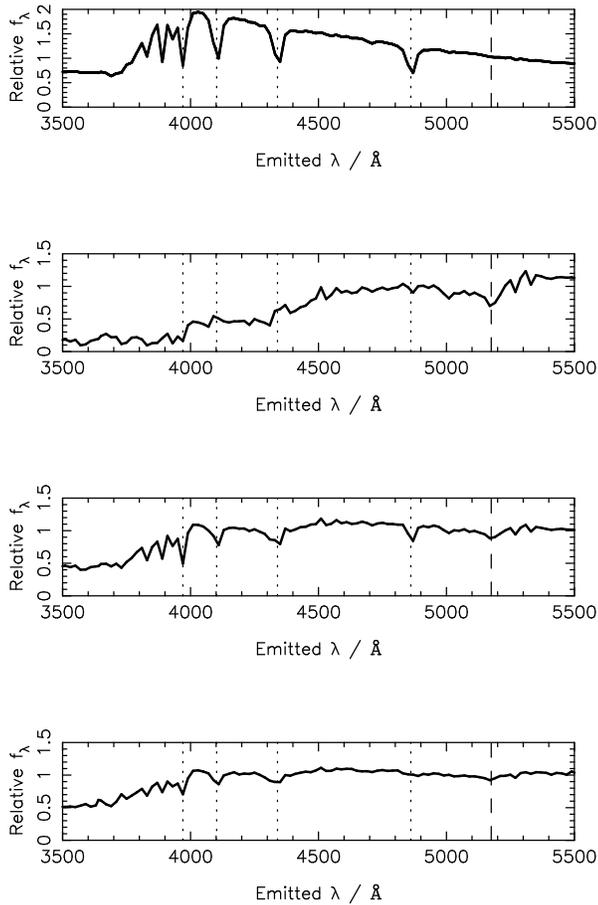}
\caption{{\it Top:} the optical 
spectrum of an A3V star; {\it second from top:} spectrum of a 12 Gyr
stellar population; {\it third from top:} combined spectrum of the A
star plus 12 Gyr stellar population: a typical E$+$A spectrum.  {\it
Bottom:} The observed spectrum of NGC 7252, from our sample, which
contains a very young $<1$~Gyr stellar population. }
\label{e+a}
\end{center}
\end{figure}

In this paper we have developed a rapid, automated method which allows
us to identify those early-type galaxies containing a significant
young ($<1$~Gyr old) population; these galaxies are E$+$A
galaxies. Fig.~\ref{e+a} shows a typical E$+$A galaxy spectrum,
together with its constituent population spectra. Unsurprisingly,
given the diagnostic abilities of our components analyses, the
component spectra look very similar to the spectra recovered from the
components analyses.

\citet{fb04} have 
analysed the IUE UV spectra of a sample of
normal galaxies using PCA. One of their components was associated with
a young stellar population, and another with a mature population,
consistent with our results. However, they were unable to correlate
the principal components with the optical morphology of the
galaxies. As we have prior knowledge of the component stellar
populations in our galaxy sample, we have been able to make a more
quantitative interpretation of our results: if the value of $a_1$ (the
weight of the `young' component) for a particular early-type galaxy is
$>$ 0.5, there is a significant $<1$ Gyr-old stellar population
present in that galaxy.

Hence, our components analysis is able to rapidly identify
post-starburst galaxies, without the uncertainties inherent in the
measurement of the equivalent widths of the Balmer absorption and
[OII] emission lines. [OII] emission in disk-dominated galaxies may
suffer from dust obscuration, and H$\gamma$ and H$\beta$ are subject
to emission-filling. These, together with the high signal-to-noise
requirements, introduce significant uncertainties in the measurement
of these lines. In contrast, as our method utilises all the
information present in the long-baseline spectra i.e. both the
features present and the overall shape of the spectra, it is a more
robust diagnostic of recent star-forming activity.

Using the wealth of data available in the SDSS and the 2dFGRS, with
signal-to-noise as good or better than the data we analyse here, we
may map precisely in what environments star-formation is suppressed
(e.g. in groups or clusters), by rapidly and robustly identifying
those galaxies hosting only mature ($>$1 Gyr) stellar
populations. This allows us to investigate whether galaxy
transformation (from star-forming to passive galaxies) occurs via
e.g. ram-pressure stripping, tidal stripping, harassment or suppression
of galaxy-galaxy interactions. Although the wavelength ranges of the
SDSS and 2dF spectra are shorter than those we have used, we can use a
training set of data to identify the linear components (as suggested
below), which can then be applied to the SDSS and 2dF data.

With the higher-resolution spectra of these surveys, and higher-resolution model spectra that are available (e.g. \citep{bc03}), we expect the accuracy of our statistical estimates to increase, as we have more independent samples (wavelength bins) of the mixed components. Whether or not the number of components, $K$, would increase would depend on the data set used. Both the existence of additional components and the interpretation of any further components can only be explored by carrying out the experiment. However, we would not expect the increase in information that we obtain from higher-resolution spectra to lead to a decrease in $K$.

We intend to extend our Bayesian modelling technique to identify other
classes of galaxies, by using appropriately selected training sets of
galaxy spectra and wavelength ranges. For example, starburst galaxies
may be identified by the Wolf-Rayet `bump' around $\sim$4600--4900
{\AA}, which is a characteristic feature of the emission from the
Wolf-Rayet stars in starburst galaxies. By first applying the Bayesian
analysis to an appropriate training set of spectra, which include the
spectral range of this feature, identification of these galaxies in a
large sample should be possible.

\section{Conclusions}

Using rigorous Bayesian modelling, we have successfully developed a
time-efficient method for the analysis of the spectra of early-type
galaxies. At $\sim$a few seconds per galaxy for this analysis, its 
speed compares very well with work by other authors \citep[\eg][]{hjl00,f05}. 
By comparing the results of 
this analysis with the results of our unique two-component synthetic 
stellar population fitting to the long-baseline data, and with two-component 
synthetic spectra, we have shown that the independent components analysis 
is physically interpretable. 

Two linear components are sufficient to describe the
bulk of the stellar content of early-type galaxies. One of the
components represents a young ($<$1 Gyr) stellar population, and the
other an old, metal-rich population. The relative contribution of
these two components allows us to robustly identify which early-type
galaxies host significant young stellar populations i.e. which are
post-starburst galaxies. 

Of the four variational Bayesian methods investigated here, We find
the variational Bayesian rectified factor analysis to be the most
useful, since it has the greatest flexibility in the shape of the
allowed distribution, whilst retaining the astronomically motivated
requirement that the flux values in each component spectrum are
non-negative.  This method achieved the highest evidence and the
highest correlation between the component weights and the population
parameters determined from spectral fitting. The method allows the
observational errors to be included in the formulation, and has the
additional powerful attribute that it can be used to recover
missing regions of spectral coverage. The strength of the correlation
between the posterior mean predictions of the data model, and the
values imputed from the best-fitting physical model, is a strong
indication that the data-driven analysis and the physical analysis are
well matched.

We intend to apply our analysis to the early-type galaxies, in the 2dF,
6dF and SDSS catalogues, as a useful tool to investigate their
evolution with environment and redshift. We have shown that
in addition to efficiently
finding early-type galaxies with significant young stellar
populations, this method would also be an efficient tool to find 
other distinctive sub-classes, such as E$+$A
galaxies, which is an important missing link in current
population scenarios of the formation of early-type galaxies.

In this work, we use model and observed spectra between
2500-8000 \AA. However, to be useful as a data mining tool for
$z\!\sim\! 0$ galaxies in the SDSS, 2dF and 6dF archives, the Bayesian
analysis has to work efficiently in the 3700--8000 \AA\ range equally
well. Our next step will be to estimate data models that will learn, from
the current datasets that include UV spectra, to work on much larger, 
solely optical datasets. 

In future studies, we therefore intend to:
\begin{itemize}
\item estimate the data models for the 3700$-$8000 \AA\ range;
\item apply our technique to 2dF, 6dF, SDSS catalogued spectra, to identify E$+$A galaxies and their environments;
\item experiment with identifying starburst galaxies;
\item consider the robustness of our data models with respect to uncertainties in the calibration and response of the 2dF/6dF/SDSS fibres, and the effect of random continuum errors, via analyses of simulated data with additional noise factors;
\item extend the models to deal with stochastic censoring, i.e. non-ignorable (versus random) missing data.
\end{itemize}

\section*{Acknowledgements}
Markus Harva was supported by a Paul \& Yuanbi Ramsay research award
at the School of Computer Science of the University of Birmingham.

\onecolumn

\appendix
\section{Implementation details of the VB-RFA data model}
\subsection{Data model estimation}
\label{updates}
\subsubsection{Posterior updates for $q(s_{kt})$}
The following simplifying notations will be used in this subsection:
$s \equiv s_{kt}, m_s\equiv\langle m_{s_k} \rangle , v_s \equiv \langle e^{v_{s_k}}
\rangle ^{-1}$. 
Further, the following expressions, involving variables from the Markov blanket (MacKay 2003) 
of a variable $s$, will be required: $v_x \equiv \ny \sum_{n=1}^N \langle e^{v_{x_n}}
\rangle  \langle a_{nk}^2\rangle \zr^{-1}$ and  
$x \equiv v_x \ny \sum_{n=1}^N \langle e^{v_{x_n}}\rangle \langle a_{nk}\rangle  \left( \langle 
x_{nt}
\rangle  - \sum_{j\neq k} \langle a_{nj}
\rangle \langle f(s_{jt})
\rangle  \right) \zr$.
Using these notations, as well as expressing $f(s)=u(s)s$, where $u(s)$ is the step 
function, then the outline of the computation of the free form posterior $q(s)$ and its 
sufficient statistics can be given as follows: 
\begin{align}
q(s) &= \frac{1}{Z} \cc(x|u(s)s,v_x)\cc(s|m_s,v_s)\\
  &= \frac{1}{Z} w_{u(s)} \cc(s|\bar{s}_{u(s)},\tilde{s}_{u(s)})
\end{align}
where 
\begin{align}
\tilde{s}_{u(s)} &= \ny u(s)/v_x + 1/v_s \zr^{-1}\\
\bar{s}_{u(s)} &= \tilde{s}\ny u(s)x/v_x + m_s/v_s \zr \\
w_{u(s)} &= \cc(x|u(s)m_s, u(s)v_s + v_x)\\
Z &= \int_{-\infty}^{\infty} ds w_{u(s)} \cc(s|\bar{s}_{u(s)},\tilde{s}_{u(s)}) \nonumber 
\\
&= \sum_{u=0}^1 w_u \ee\ny (-1)^u \bar{s}_u/\sqrt{2\tilde{s}_u} \zr \label{Z}.
\end{align}
For more detailed derivation, see \citet{harva05}.

Note the non-trivial variational posterior (A2) that resulted from the free-form approximation, a bimodal distribution (A6), which is essentially a mixture of two rectified Gaussians.

The posterior statistics associated with $q(s)$ that are required for updating the posterior statistics of other variables 
of the model are the first and second order moments, as follows. Denoting $e_u=\ee\ny (-1)^u \bar{s}_u/\sqrt{2\tilde{s}_{u}} \zr$ (which is the $\ee$-term in (\ref{Z})), 
the $i$-th order moments of $s$ and $f(s)$ can be written as
\begin{equation}
\langle f^i(s) \rangle = \int f^i(s)q(s)ds = M_i(u=1),
\end{equation}
where 
\begin{equation*}
M_i(u)\equiv\frac{w_u}{2Z} e_u  \int s^i \cc^R(s|(-1)^{1-u}\bar{s}_u,\tilde{s}_u)\, ds,
\nonumber
\end{equation*}
and
\begin{equation}
\langle s^i
\rangle  = \sum_{u=0}^1 M_i(u).
\end{equation}
The notation $M_i(u)$ introduced in (A7) will be used also in
(\ref{qs}). The remaining integral is the $i$-th moment of a rectified
Gaussian density. The exact expressions for these, are given in \S
\ref{rectG} and more details on it can be found e.g. in
\citet{harva04}, 
along with numerical issues concerning an efficient implementation.

\subsubsection{Posterior updates for $q(m_{s_k})$}
The posteriors $q(m_{s_k})=\cc
(m_{s_k}|\bar{m}_{s_k},\tilde{m}_{s_k})$ are of a Gaussian form, where
\begin{align*}
\tilde{m}_{s_k} &= \ny \lln e^{v_{m_{s}}} \rrn + T\langle e^{v_{s}}
\rangle  \zr^{-1}\\
\bar{m}_{s_k} &= \tilde{m}_{s}\left( \lln e^{v_{m_{s}}}\rrn \lln m_{m_{s}}\rrn +\langle 
e^{v_{s_k}}
\rangle \sum_{t=1}^T \langle s_{kt}
\rangle  \right)
\end{align*}
which provide the required posterior expectation 
$\langle m_{s_k}
\rangle  = \bar{m}_{s_k}$. 
Note that if a `vague prior' was used on $m_{s_k}$, then the posterior mean estimate $\langle m_{s_k} \rangle $ becomes similar to a maximum likelihood estimate $\sum_t \lln 
s_{kt} \rrn/T$, as then $\lln m_{m_{s}} \rrn = m_{m_{s}}=0$ and $\lln e^{v_{m_{s}}}\rrn = 
e^{v_{m_{s}}}\approx 0$.

\subsubsection{Posterior updates for $q(a_{nk})$}
\label{qx}
The posteriors $q(a_{nk})=\cc^R(a_{nk}|\bar{a}_{nk},\tilde{a}_{nk})$ are rectified 
Gaussians, where
\begin{align*}
\tilde{a}_{nk} &= \ny 1+\langle e^{v_{x_n}}
\rangle \sum_{t=1}^T \langle f^2(s_{kt})
\rangle  \zr^{-1}\\
\bar{a}_{nk} &= \tilde{a}_{nk}\langle e^{v_{x_n}}
\rangle \sum_{t=1}^T \left( \langle x_{nt}
\rangle  - \sum_{j\neq k} \langle a_{nj}
\rangle \langle f(s_{jt})
\rangle  \right) \langle f(s_{kt})
\rangle 
\end{align*}
which provide the required sufficient statistics: 
$\langle a_{nk}^i
\rangle  = \int a_{nk}^i \cc^R (a_{nk}|\bar{a}_{nk}, \tilde{a}_{nk}) da_{nk}$. 

\subsubsection{Posterior updates for $q(x_{nt})$}
The posteriors $q(x_{nt})=\cc(x_{nt}|\bar{x}_{nt},\tilde{x}_{nt})$ are Gaussian, where
\begin{align*}
\tilde{x}_{nt} &=\ny \sigma^{-2}_{y_{nt}}b_{nt}+\langle e^{v_{x_n}}
\rangle  \zr^{-1}\\
\bar{x}_{nt} &= \tilde{x}_{nt} \left( \sigma^{-2}_{y_{nt}}b_{nt}y_{nt} + \langle 
e^{v_{x_n}}
\rangle \sum_{k=1}^K \langle a_{nk}
\rangle \langle f(s_{kt})
\rangle \right)
\end{align*}
These provide the required posterior expectation $
\langle x_{nt}
\rangle  = \bar{x}_{nt}$. 
Here, the indicator variable $b_{nt}=1$ if $y_{nt}$ is observed and zero if it is missing.

\subsubsection{Posterior updates for variance parameters}
Following \citet{valpola05} and \citet{vhk04}, in the
$\exp$-parameterisation (see \S4.1.1), Gaussian priors are imposed on
$v_{\theta}$, where $\theta$ stands for any of the variables $x$, $s$
or $m$ in Fig.~\ref{gr_model}. This is then solved by a fixed form
Gaussian variational posterior,
\begin{equation}
q(v_{\theta})=\cc (v_{\theta}|\bar{v}_{\theta},\tilde{v}_{\theta}),
\end{equation}
and this has been employed in the experiments described in this paper.

Here, $\bar{v}_{\theta}$ and $\tilde{v}_{\theta}$ are computed by
numerical optimisation \citep{valpola05,vhk04}.  Alternatively, a
direct parameterisation may be used with Gamma priors and Gamma
posteriors may be sufficient for this model.

\subsection{The log evidence bound}
\label{evidence_eq}
Computing the log of the evidence bound is useful for monitoring the
convergence of the iterative estimation process, and more importantly,
for comparing different modelling hypotheses.
\label{evidence}
As mentioned, (\ref{evid}) decomposes into a number of terms.
\begin{equation}
\log p(\bd{Y}) \geq \sum_{nt}\langle \log p(y_{nt}|x_{nt}) \rangle
+\sum_{\theta \in \bd{\theta}} \ny \langle \log p(\theta|\pa(\theta)) \rangle -  \langle \log q(\theta) \rangle \zr .
\end{equation}
The $p$ and $q$ terms for Gaussians will be given only once, while those that are 
different will be detailed separately.

\subsubsection{$p$- and $q$-terms of Gaussian variables}
The $p$ and $q$ terms in (A10), in the case of Gaussian variables, can be written as:
\begin{align*}
\langle \log p(y_{nt}|x_{nt},\sigma_{nt}^2)\rangle 
 &= - \frac{1}{2} 
\ny \frac{1}{\sigma_{nt}^2}\left[
        (y_{nt} -  \lln x_{nt}\rrn)^2 + \Var{x_{nt}} \right]
     + \log 2\pi\sigma_{nt}^2
\zr \\
\langle \log p(\theta|m_{\theta},e^{-v_{\theta}})\rangle 
 &= - \frac{1}{2} 
\ny 
  \lln e^{v_{\theta}}\rrn \left[
    (\bar{\theta} - \lln m_\theta \rrn)^2 + \tilde{\theta} + \Var{m_\theta}
  \right]
  - \lln v_\theta \rrn + \log 2\pi
\zr \\
- \lln \log q(\theta) \rrn &= \frac{1}{2}+\frac{1}{2}\log 2\pi\tilde{\theta},
\end{align*}
where $\bar{\theta}$ and $\tilde{\theta}$ are the posterior mean and variance of $\theta$. 
If the $\exp$-parameterisation is used for the variances, then 
the term 
$\lln e^{v_{\theta}} \rrn = e^{\bar{v}_\theta + \tilde{v}_{\theta}/2}$ 
 --- see \cite{vhk04}.

\subsubsection{$p$- and $q$-terms of $a_{nk}$}
The $p$ and $q$ terms of $a_{nk}$, which are the elements of the matrix $\bd{A}$ in (A1), are given by:
\begin{align*}
\langle \log \cc^R(a|0,1)\rangle 
 &= - \frac{1}{2} 
\ny \log 2\pi + \lln a\rrn^2 + \Var{a} \zr + \log 2\\
- \lln \log q(a) \rrn 
 &= \frac{1}{2\tilde{a}} \ny \Var{a} + (\lln a \rrn - \bar{a})^2\zr 
- \frac{1}{2}\log\frac{2}{\pi\tilde{a}}  
+ \log \ee (-\bar{a}/\sqrt{2\tilde{a}}),
\end{align*}
where we have dropped the indices.

\subsubsection{The $q$-term of $s_{kt}$}
The $p$-term of variable $s$ is computed as in the case of other Gaussian variables. The $q$-term gives the following:
\begin{equation}
\label{qs}
 -\lln \log q(s) \rrn = \sum_{u=0}^1 
  \ny \left[\frac{\bar{s}_u^2}{2 \tilde{s}}  
             - \log \frac{w_u}{Z\sqrt{2\pi\tilde{s}_u}}\right] M_0(u)
    - \frac{\bar{s}_u}{\tilde{s}_u}M_1(u) + \frac{1}{2\tilde{s}_u} M_2(u)
  \zr .
\end{equation}
For detailed derivation see \cite{harva05}.

\subsection{Missing value prediction}
\label{miss}
A Gaussian approximation of the predictive distribution can be
obtained as follows.
\begin{align}
q(y_{nt}^m|\bd{y}_t^o) &= \int d\bd{\theta}p(y_{nt}^m|\bd{\theta})q(\bd{\theta}) \\
&= \int d\bd{\theta}_{-x_{nt}} \int 
dx_{nt}p(y_{nt}|x_{nt})p(x_{nt}|\bd{\theta}_{-x_{nt}})q(\bd{\theta}_{-x_{nt}}) \\
&= \int d\bd{\theta}_{-x_{nt}} \cc (y_{nt}^m|\sum_k a_{nk}f(s_{kt}), 
\sigma^2_{y_{nt}^m}+e^{-v_{x_n}})q(\bd{\theta}_{-x_{nt}}) \label{xxx}\\
&\approx \cc (y_{nt}^m|\sum_k \langle a_{nk}
\rangle \langle f(s_{kt})
\rangle , \sigma^2_{y_{nt}^m}+\langle e^{v_{x_n}}
\rangle ^{-1}) \label{pred}
\end{align}
where $\bd{\theta}_{-x_{nt}}$ denotes hidden variables above $x_{nt}$ in the conditional 
dependency architecture of the model. The mean of this distribution provides a prediction 
for the missing entry $y_{nt}$. Conveniently, for missing entries the mean of (\ref{pred}) 
is identical to that of $q(x_{nt})$ (as $b_{nt}=0$), so that the imputation of missing 
values takes no additional computation. Further,
evaluating (\ref{pred}) when $\sigma_{y_{nt}}^2=0$ (i.e. no measurement of errors) -- in order to test the prediction performance -- also happens to be identical to evaluating $q(x_{nt})$ from (A15) and \S A1.4.


\subsection{Required properties of $\cc^R(\theta|\bar{\theta},\tilde{\theta})$}
\label{rectG}
The rectified Gaussian distribution along with its first and second order moments, used in \S A1.1 and \S A1.3, are listed here for completeness. 
\begin{equation}
\cc^R(\theta|\bar{\theta},\tilde{\theta})=
\frac{2}{\ee(\bar{\theta}/\sqrt{2\tilde{\theta}})}
u(\theta)\cc(\theta|\bar{\theta},\tilde{\theta})
\end{equation}
\begin{align*}
\langle \theta \rangle &= \int  \theta \cc^R(\theta|\bar{\theta},\tilde{\theta})d\theta=
\bar{\theta} + \sqrt{\frac{2\tilde{\theta}}{\pi}}
\frac{1}
{\exp\ny (\bar{\theta}/\sqrt{2\tilde{\theta}})^2 \zr 
\ee(-\bar{\theta}/\sqrt{2\tilde{\theta}})}\\
\langle \theta^2 \rangle &= \int \theta^2 
\cc^R(\theta|\bar{\theta},\tilde{\theta})d\theta=
\bar{\theta}^2 + \tilde{\theta} + \sqrt{\frac{2\tilde{\theta}}{\pi}}
\frac{\bar{\theta}}
{\exp\ny (\bar{\theta}/\sqrt{2\tilde{\theta}})^2 \zr 
\ee(-\bar{\theta}/\sqrt{2\tilde{\theta}})}
\end{align*}
See \citet{harva04} or \citet{miskin00}
for details.

\end{document}